\documentclass[pra,byrevtex,showpacs,superscriptaddress,nofootinbib,twocolumn]{revtex4-1}
\usepackage{graphicx}
\usepackage{amsmath}
\usepackage{bm}
\usepackage{amssymb}
\usepackage{mathrsfs}
\usepackage{hyperref}
\hypersetup{colorlinks=true,linkcolor=blue,citecolor=blue,urlcolor=blue}

\begin{document}

\title{Deterministic and probabilistic entanglement swapping of nonmaximally entangled states assisted by optimal quantum state discrimination.}

\author{M. A. Sol\'is-Prosser}\email{miguelangel.solis@cefop.udec.cl} 
\author{A. Delgado}
\affiliation{Center for Optics and Photonics, Universidad de Concepci\'on, Casilla 4016, Concepci\'on, Chile}
\affiliation{MSI-Nucleus on Advanced Optics, Universidad de Concepci\'on, Casilla 160-C, Concepci\'on, Chile}
\affiliation{Departamento de F\'isica, Universidad de Concepci\'on, Casilla 160-C, Concepci\'on, Chile}

\author{O. Jim\'{e}nez}
\affiliation{Departamento de F\'isica, Facultad de Ciencias B\'asicas, Universidad de Antofagasta, Casilla 170, Antofagasta, Chile}

\author{L. Neves}
\affiliation{Departamento de F\'isica, Universidade Federal de Minas Gerais,
Caixa Postal 702, Belo Horizonte, MG 30123-970, Brazil}

\date{\today}

\begin{abstract}

We analyze entanglement swapping (ES) of partially entangled pure states with arbitrary Schmidt rank from the perspective of quantum state discrimination. It is shown that the standard deterministic ES protocol is related with an optimal  minimum-error strategy. In this case the amount of entanglement of the states resulting from swapping is, in general, lower than the maximum achievable for the quantum channels involved. In this regard, we show that the ES protocol can be probabilistically improved resorting to optimal maximum-confidence (MC) discrimination strategy. Additionally, we show that the success probability of achieving entanglement above a prescribed value from standard deterministic ES can be increased by applying sequential MC measurements.
\end{abstract}

\date{\today}
\pacs{03.67.-a, 03.65.Ud, 03.67.Hk}
\maketitle

\section{Introduction}
\par At present, entanglement plays the role of a resource which allows the implementation of many quantum information processing tasks. In particular, it has been applied to quantum computation~\cite{Ladd10,Kok07}, quantum cryptography~\cite{Gisin02,Jennewein00,Naik00,Tittel00}, quantum communications~\cite{Cleve97}, and quantum teleportation~\cite{Bennett93}.

A particularly intriguing property of entangled states is the possibility of entangling quantum systems that have never interacted or never had a common past. This process, referred to as entanglement swapping (ES)~\cite{Bennett93,Zukowski93}, requires two pairs of maximally entangled systems. One system of each pair is sent to Charlie, while the two other systems are sent one to Alice and one to Bob. Afterwards, Charlie carries out a Bell-state measurement (BSM) on his pair of systems. This projects Alice's and Bob's systems onto a maximally entangled state irrespectively of Charlie's measurement result. Important applications of ES are quantum repeaters~\cite{Sangouard11}, which allow one to generate entanglement between distant users~\cite{Briegel98,Dur99,Waks02}, as an entanglement concentration scheme~\cite{Shi00,Hsu02,Modlawska08,Yang09}, which allows one to increase probabilistically the amount of entanglement between two parties, and in experimental studies on nonlocality~\cite{Zukowski93}. Further applications are generation of Greenberger-Horne-Zeilinger (GHZ) states via multiparticle ES~\cite{Hardy00,Bose98}, quantum secret sharing~\cite{Zhou09} and quantum communication protocols~\cite{Xia07,Zhan09,Qin10,Scherer11,Zhou11}.

Experimental demonstrations of ES have been performed via nuclear magnetic resonance (NMR) quantum processors~\cite{Boulant03}, trapped ions~\cite{Riebe08}, and photons in the visible wavelength range~\cite{Pan98,Jennewein01,Sciarrino02,Kaltenbaek09}. Time-bin-based ES was also demonstrated at a telecommunication wavelength of 1550 nm~\cite{Riedmatten05,Halder07,Takesue09,Sangouard11exp}. At this wavelength, polarization ES has also been reported~\cite{Xue12}. Recently, an experiment on delayed ES~\cite{Peres00} was implemented~\cite{Ma12}. Other experimentally demonstrated applications include multistage ES~\cite{Goebel08}, and generation of GHZ states~\cite{Lu09} by means of ES.

Typically, ES is formulated considering maximally entangled states. However, from the experimental point of view, maximal entanglement is difficult to generate, although highly entangled states are feasible. For this reason ES has been studied assuming partially entangled states. Initially, the case of two identical partially entangled two-qubit states was considered. In this case the entanglement generated between Alice and Bob equals the optimal amount of entanglement that could be concentrated if one of the initial states were directly distributed between Alice and Bob~\cite{Bose99}. Thereafter, this result was extended to two different partially entangled two-qubit states. In this case the entanglement generated between Alice and Bob corresponds to that which could be concentrated from the less entangled of the two states if this were distributed directly between them~\cite{Shi00}. A different approach was followed by noting the existence of a link between ES and quantum state discrimination~\cite{Delgado05}. Here it was shown that Charlie's BSM, when implemented via a generalized controlled-{\sc not} ({\sc c-not}) gate \cite{Alber01,Alber00}, generates sets of nonorthogonal bipartite separable states. Consequently, these states cannot be deterministically discriminated. Nevertheless, it is possible to resort to a particular protocol for quantum state discrimination, namely unambiguous discrimination (UD) \cite{Ivanovic87,Dieks88,Peres88,Jaeger95,Peres98}. This scheme allows one to identify nonorthogonal states with certainty but at the expense of introducing an inconclusive outcome which occurs at a rate that can be minimized. Thereby, it is possible to distinguish the states generated by Charlie and to generate a maximally entangled state between Alice and Bob at an optimal discrimination rate which agrees with the optimal concentration rate and the previously obtained results~\cite{Bose99,Shi00}.

The exploitation of the link between ES and state discrimination was limited to a very small class of pairs of partially entangled states. This limitation arises because UD exists only for sets of linearly independent states \cite{Chefles98}, which restricts strongly the states initially shared by Alice, Bob and Charlie to carry out ES. Here we remove this restriction and present a complete analysis of ES when implemented via a pair of different partially entangled states of $D$-dimensional quantum systems (qudits). Both states have, in general, different Schmidt rank. In this case, the generalized {\sc c-not} gate involved in the BSM generates several sets of nonorthogonal states. Some of these sets can be linearly dependent and thus these are not well suited for UD. However, it has been shown that linearly dependent states can also be discriminated via maximum confidence (MC) \cite{Croke06}. In this discrimination strategy it is necessary to give up the certainty when identifying the nonorthogonal states and, consequently, there exists only a certain confidence degree that the measurement results indicate a particular nonorthogonal state. Such a confidence degree can be maximized and achieved with an optimal success probability. In general, MC corresponds to an optimization problem and, consequently, it lacks of general solutions. Fortunately, in the case of ES the states generated by the BSM, through the generalized {\sc c-not} gate, turn out to be symmetric and with equal \emph{a priori} probabilities, for which analytical solutions in the case of MC are available \cite{Jimenez11,Herzog12}. Thereby, it is possible to design a ES protocol assisted by MC (MCaES). This protocol replaces the BSM by the concatenation of a generalized {\sc c-not} gate to a stage of MC discrimination and can be applied to any pair of partially entangled two-qudit states without restrictions on their rank or Schmidt coefficients.  We show that conclusive events in MCaES lead to a probabilistic increase in entanglement and fidelity when compared to the standard ES protocol. Furthermore, due to some features of MC it is possible to reuse the postmeasurement states arising from a failure in the discrimination attempt. In this case the postmeasurement states can also be discriminated in a second MC stage, which contributes to increase the possibility to achieve an entanglement or fidelity above a prescribed value. This procedure can be iterated several times depending on the specific values of the Schmidt coefficients of the states involved in the ES process. We present numerical simulations which show that, depending on the Schmidt coefficients of the states owned by Alice and Bob at the beginning of the protocol and on the outcomes obtained at other measurements, one or two additional discriminations after a failed first attempt can still offer better values for some figures of merit than by applying the standard process. In consequence, the implementation of a multistage MCaES increases the probability to outperform the standard ES.

This article is organized as follows. We start by briefly describing an ideal standard ES protocol between maximally entangled qudits in Sec.~\ref{sec:standardES}. Then, in Sec.~\ref{sec:noMES}, we discuss the case of ES using two pairs of arbitrary entangled qudits and its relationship with quantum state discrimination among nonorthogonal states. This problem is then addressed in Secs.~\ref{sec:ME} and~\ref{sec:MC}, comparing the performance of ES in connection to minimum-error (ME) and MC, respectively. Section~\ref{sec:SMC} introduces the usage of sequential MC measurements applied to enhance the ES protocol and Sec.~\ref{numerical} shows numerical simulations aimed to understand differences between the different methods. Finally, Sec.~\ref{sec:Summary} concludes this work.

\section{Standard entanglement swapping protocol\label{sec:standardES}}

\par
\begin{figure}[t]
	\centering
	\includegraphics[width=0.45\textwidth]{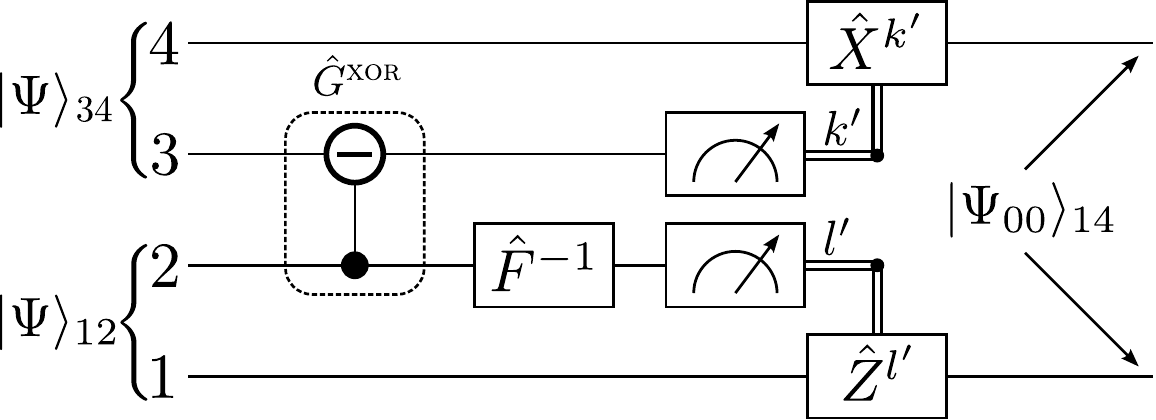}
	\caption{Quantum circuit for standard ES protocol. When $|\Psi\rangle_{12}$ and $|\Psi\rangle_{34}$ are maximally entangled states, this process always leads to obtain a maximally entangled state. Conditional unitary corrections can be made by the parties to share any particular Bell state, as Eq.~(\ref{bellstates}) indicates.}
	\label{fig:circuit1}
\end{figure}

 Let us consider two pairs of particles labeled by 1-2 and 3-4. Each pair is owned by different parties usually named Alice and Bob. These particles will be described by two maximally entangled states, each one belonging to a $D\times D$-dimensional Hilbert space, as
\begin{align}
	|\Psi\rangle_{1234} = \frac{1}{\sqrt{D}}\sum_{m=0}^{D-1}|m\rangle_{1}|m\rangle_{2}\otimes\frac{1}{\sqrt{D}}\sum_{n=0}^{D-1}|n\rangle_{3}|n\rangle_{4},
\end{align}
where each bipartite state was written in its Schmidt basis. We shall assume that the Schmidt basis coincides with the computational basis for each system. Now, a third party named Charlie takes one particle from Alice and Bob and performs a BSM on them. This can be done by applying a $\hat{G}^\text{\sc xor}$ operation followed by an inverse Fourier transform on particle $2$ and projective measurements on the computational basis for particles $2$ and $3$~\cite{Brassard98,NielsenBook}. The $\hat{G}^\text{\sc xor}$ gate is defined by~\cite{Alber01}
\begin{align}
	\hat{G}^\text{\sc xor}_{23}|m\rangle_{2}|n\rangle_{3} = |m\rangle_{2}|m \ominus n\rangle_{3},
\end{align}
where, in this case, particles 2 and 3 have been chosen as control and target, respectively. The symbol $\ominus $ denotes subtraction modulo $D$. It can be shown that by applying the $\hat{G}^\text{\sc xor}_{23}$ gate allows one to write the state of the whole set of particles as
\begin{align}
	\hat{G}^\text{\sc xor}_{23}|\Psi\rangle_{1234} &= \frac{1}{D}\sum_{l,u=0}^{D-1}|\Psi_{lu}\rangle_{14} ~ \hat{F}_{2}|l\rangle_{2} ~ |u\rangle_{3},\label{xor_max}
\end{align}
where $\hat{F}_{2}^{}$ is the Fourier transform operator acting on system 2, 
\begin{align}
	\hat{F}=\frac{1}{\sqrt{D}}\sum_{n,q=0}^{D-1}e^{2\pi inq/D}|n\rangle\langle q|,\label{FourierTransformD}
\end{align}
and the bipartite states $|\Psi_{lu}\rangle_{14}$ are a generalization of the Bell states for qudits. These entangled states are defined as~\cite{Alber00}
\begin{align}\nonumber
	|\Psi_{lu}\rangle_{14} &= \frac{1}{\sqrt{D}}\sum_{q=0}^{D-1}e^{-2\pi i ql/D}|q\rangle_{1}|q \ominus  u\rangle_{4} \\
	&=\hat{Z}_{1}^{-l}\hat{X}_{4}^{-u}|\Psi_{00}\rangle_{14},\label{bellstates}
\end{align}
where $\hat{Z}$ and $\hat{X}$ are the phase and shift operators that operate, on the computational basis, as $\hat{Z}|l\rangle=e^{2\pi i l/D}|l\rangle$ and $\hat{X}|l\rangle=|l\oplus 1\rangle$, respectively. From Eq.~(\ref{xor_max}) it is clear that Charlie can operate with an inverse Fourier transform on particle 2 and then measure both particles 2 and 3 on the computational basis. Once this has been done, the state of the particles 1 and 4 will be $|\Psi_{l'u'}\rangle_{14}$ after Charlie communicates, via a classical channel, his results to Alice and Bob. Therefore, it is possible to transform the state of particles 1 and 4 into a maximally entangled state even though they were not initially entangled and never interacted directly. Alice and Bob might want to share a specific $|\Psi_{lu}\rangle$ state as the initial states owned by them. In this case, the parties can apply a phase and shift operations on their particles conditional to the measurements outcomes obtained by Charlie. The circuit of Fig.~\ref{fig:circuit1} shows an example of the full procedure for this goal when Alice and Bob want to share the $|\Psi_{00}\rangle$ state.

\section{Entanglement swapping with nonmaximally entangled states \label{sec:noMES}}
\par It has been shown~\cite{Delgado05} that ES using two pairs of $D\times D$-dimensional nonmaximally entangled states is intimately connected to the problem of discriminating among $D$ linearly independent states. This work considered two pairs of different nonmaximally entangled states whose Schmidt rank is maximal and belong to Hilbert spaces that have the same dimension $D$. In the present work we will consider two pairs of bipartite entangled states lying in Hilbert spaces whose dimensions, in general, differ. Their Schmidt ranks are arbitrary and, as well as the Schmidt coefficients themselves, can differ. Thus, let $|\Psi\rangle = |\phi\rangle_{A}|\varphi\rangle_{B}$ be the state of this four-particle system, where $|\phi\rangle_{A}$ and $|\varphi\rangle_{B}$ are written in their Schmidt decomposition as 
\begin{align}
	|\phi\rangle_{A} &= \sum_{m=0}^{D_{A}-1}c_{m}|m\rangle_{1}|m\rangle_{2},\\
	|\varphi\rangle_{B} &= \sum_{r=0}^{D_{B}-1}d_{r}|r\rangle_{3}|r\rangle_{4}.
\end{align}
The number $D_{A}$ ($D_{B}$) denotes the dimension of the Hilbert space spanned by the Schmidt basis of one of the particles of Alice (Bob). We impose to the coefficients $c_{m}$ and $d_{n}$ to be real non-negative and normalized. As well as it was assumed in the previous section, the computational basis coincides with the Schmidt basis. Following the standard procedure for ES described above, Charlie will take particles 2 and 3 to operate with a $\hat{G}^\text{\sc xor}$ gate on them. Although the basis $\{|m\rangle_{2}\}$ and $\{|r\rangle_{3}\}$ could have different cardinality, the $\hat{G}^\text{\sc xor}$ gate can be still defined. As it was pointed out by Daboul {\it et al.}~\cite{Daboul03}, it is possible to define a hybrid\footnote{A \emph{hybrid} gate is understood as a bipartite gate in which the the dimension of the control and target Hilbert spaces can differ~\cite{Daboul03}.} version of the $\hat{G}^\text{\sc xor}$ gate for bipartite systems as $\hat{G}^\text{\sc xor}|m\rangle_{C}|r\rangle_{T}=|m\rangle_{C}|m\ominus r\rangle_{T}$, where $0\leqslant m\leqslant (D_C-1)$, $0\leqslant r\leqslant (D_T-1)$, $D_C$ ($D_T$) denotes the Hilbert space dimension of the control (target) system, and $\ominus $ denotes subtraction modulo $D_T$. Thereby, when Charlie applies this gate on particles 2 and 3, the state of the whole system becomes
\begin{align}
	\hat{G}^\text{\sc xor}_{23}|\phi\rangle_{A}|\varphi\rangle_{B} &= \frac{1}{\sqrt{D_{A}}}\sum_{l=0}^{D_{A}-1}\sum_{s=0}^{D_{B}-1}\sqrt{p_{s}}|\Psi_{ls}\rangle_{14}|\nu_{ls}\rangle_{2}|s\rangle_{3},\label{xorphiphi}
\end{align}
where $	p_{s}=\sum_{m=0}^{D_{A}-1}c_{m}^{2}d_{m\ominus s}^{2}$ is the probability of finding the third particle in the state $|s\rangle_{3}$, and $|\Psi_{ls}\rangle_{14}$ and $|\nu_{ls}\rangle_{2}$ are given by 
\begin{align}
	|\Psi_{ls}\rangle_{14} &= \frac{1}{\sqrt{D_{A}}}\sum_{q=0}^{D_{A}-1}\omega^{-ql}|q\rangle_{1}|q\ominus s\rangle_{4},\label{Psistates}\\
	|\nu_{ls}\rangle_{2} &= \sum_{m=0}^{D_{A}-1}\Gamma_{ms}\omega^{ml}|m\rangle_{2} = \hat{Z}_{2}^{l}|\nu_{0s}\rangle_{2},\label{nustates}
\end{align}
with ${\omega=\exp(2\pi i/D_A)}$, ${\Gamma_{ms}=c_{m}d_{m\ominus s}/\sqrt{p_{s}}}$, and ${\hat{Z}=\sum_{n=0}^{D_{A}-1}\omega^{ n}|n\rangle\langle n|}$. 
Since the states are normalized, the constraint ${\sum_{q=0}^{D_A -1}\Gamma_{qs}^2=1}$ is satisfied. The bipartite states $|\Psi_{ls}\rangle_{14}$ form an orthonormal basis for $\mathcal{H}_{14}=\mathcal{H}_1\otimes\mathcal{H}_4$ and, when $D_B\geqslant D_A$, represent maximally entangled states within subspaces of $\mathcal{H}_{14}$ whose dimension is $D_{\rm min}\times D_{\rm min}$, where ${D_{\rm min}=\min(D_A,D_B)}$. We shall not study the case $D_B<D_A$ along this paper since this case yields states that do not represent the most entangled states within their subspaces. 

\par Once Charlie has carried out the $\hat{G}^\text{\sc xor}$ operation, he can measure particle 3 on its Schmidt basis. As a result, the state of the remaining particles will be given by 
\begin{align}
	|\Xi_{s}\rangle_{124}= \frac{1}{\sqrt{D_{A}}}\sum_{l=0}^{D_{A}-1}|\Psi_{ls}\rangle_{14}|\nu_{ls}\rangle_{2},\label{Xis}
\end{align}
if the outcome of the measurement is $s$, which happens with probability $p_s$. To accomplish the ES protocol, Charlie must perform a measurement on particle 2 and communicate the result to Alice and Bob in order to determine the state of particles 1 and 4. But the nonorthogonality of the states $|\nu_{ls}\rangle$ is a drawback for this measurement given that they cannot be always identified with certainty. So, the problem of ES has been reduced to discriminate among the states within a set $\Omega_{s}=\{|\nu_{ls}\rangle, l=0,\dots,D_{A}-1\}$, which are symmetric and equally likely.\footnote{A set of $N$ states $\{|\psi_j\rangle,\,j=0,\dots,N-1\}$ is \emph{symmetric} under the action of a unitary operator $\hat{W}$ if they satisfy the following conditions: $|\psi_j\rangle = \hat{W}|\psi_{j-1}\rangle = \hat{W}^j |\psi_0\rangle$, and $\hat{W}|\psi_{N-1}\rangle=|\psi_0\rangle$. From Eq.~(\ref{nustates}) we can note that the states $|\nu_{ls}\rangle$, for any fixed value of $s$, form a set of $D_A$ symmetric states under the action of the operator $\hat{Z}$. These states are equally likely because the reduced density operator of system 2---from the state of Eq.~(\ref{Xis})---is $\hat{\rho}_{s} = \frac{1}{D_{A}}\sum_{l=0}^{D_{A}-1} |\nu_{ls}\rangle\langle\nu_{ls}|$, having each state the same \emph{a priori} probability $1/D_A$.} Since $s=0,\dots,D_{B}-1$, there exist $D_{B}$ of these sets. The performance of the protocol will depend on which discrimination strategy Charlie will adopt. This choice depends whether the states to be discriminated are linearly independent. The linear dependence of these states can be determined through the determinant of their Gram matrix $[\mathcal{G}_{s}]_{jk}= {}_{2}\langle\nu_{js}|\nu_{ks}\rangle_{2}$. From Eq.~(\ref{nustates}), we have that
\begin{align}
	[\mathcal{G}_{s}]_{jk} &=\sum_{l=0}^{D_{A}-1}\Gamma_{ls}^{2}\omega^{l(k-j)}.\label{Gjk}
\end{align}
Note that $\mathcal{G}_{s}$ is circulant since, from Eq.~(\ref{Gjk}), $[\mathcal{G}_{s}]_{jk} = [\mathcal{G}_{s}]_{j+1,k+1}$. In consequence, this matrix can be turned diagonal, without disturbing its determinant, by applying a discrete inverse Fourier transform matrix $\hat{F}^{-1}$ on it. Thus, we obtain

\begin{equation}
	\det(\mathcal{G}_{s}) = \det({F}^{-1}\mathcal{G}_{s}{F})
	= D_{A}^{D_A} \times \prod_{j=0}^{D_{A}-1} \Gamma_{js}^{2}.
\end{equation} 
This determinant is zero if and only if the states used to calculate it are linearly dependent. Therefore, the states of a given set $\Omega_{s}$ are linearly independent if there is any vanishing combination $c_{j}^{2}d_{j\ominus s}^{2}$. As an illustrative example, if $D_{A}=2$, $D_{B}=3$ and the only null coefficient is $d_{2}$, we will have that ${\det(\mathcal{G}_{0})\propto c_{0}^{2}d_{0}^{2}c_{1}^{2}d_{1}^{2}\ne 0}$, showing that $\Omega_0$ is a set of linearly independent states, and ${\det(\mathcal{G}_{1})\propto c_{0}^{2}d_{1}^{2}c_{1}^{2}d_{2}^{2} =0 = \det(\mathcal{G}_{2})}$, indicating that both $\Omega_1$ and $\Omega_2$ contain linearly dependent states. This example shows that for certain pairs of initially entangled states $|\phi\rangle_{A}$ and $|\varphi\rangle_{B}$ we can obtain sets of both linearly independent and linearly dependent states conditional to the obtained outcome of the measurement on particle 3. An interesting point can be discussed here: The bipartite system of the aforementioned example can also be considered as an effective $D_A'\times D_B'=2\times 2$ system. In such a case, two sets $\Omega_0'$ and $\Omega_1'$ are obtained, containing two linearly independent states each one. A natural question that arises is about the difference between these two approaches. The answer can be found by looking at the space where the $\hat{G}_{23}^\text{\sc xor}$ gate acts on. If it is possible for Charlie to adapt his quantum gate to make it act only on the \emph{effective} bipartite Hilbert space (in this case, the $2\times 2$-dimensional subspace), we can consider the latter approach. Otherwise, if his gate has been predesigned to act on arbitrary vectors lying on a $D_A\times D_B$-dimensional bipartite space, the first approach should be considered.

Recapitulating, Eq.~(\ref{Xis}) shows that the ability of discriminating among a set of nonorthogonal states in an optimal way is essential to carry out an optimal ES protocol when the initial entangled states are not maximally entangled. Many discrimination strategies have been developed so far~\cite{BergouReview,BergouBook} and the one Charlie will apply depends on which type of ES protocol he, together with Alice and Bob, agreed to implement beforehand. In the following, we shall analyze two protocols. First, we consider that ES is implemented deterministically. In this case, we show that Charlie must apply ME strategy. Second, we consider a probabilistic protocol where the goals are to maximize the entanglement of the final state and the success probability of doing so. In this case, we show that Charlie must apply optimized MC measurements.

\subsection{ES assisted by ME discrimination \label{sec:ME}}

\begin{figure}[!t]
	\centering
	\includegraphics[width=0.45\textwidth]{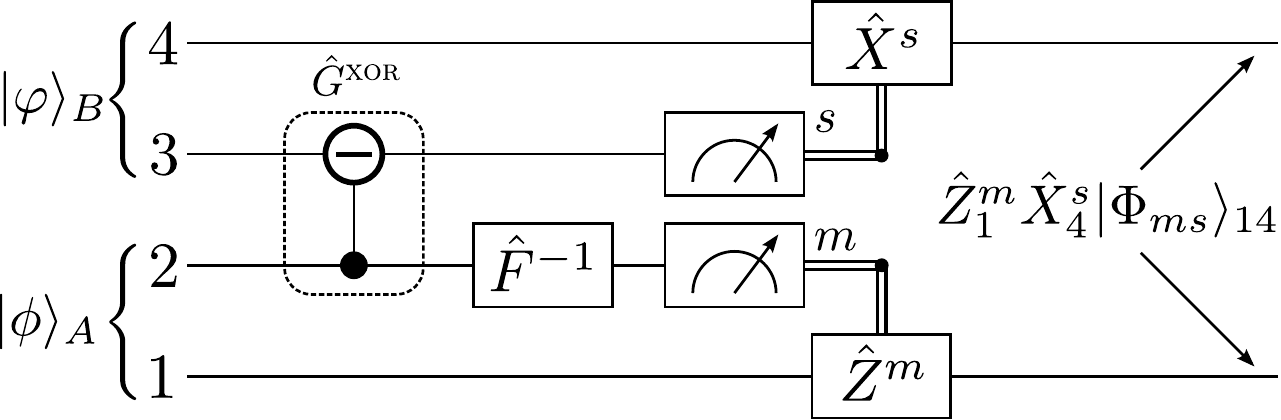}
	\caption{Quantum circuit for ES protocol assisted by ME discrimination. The operations remain the same ones of the standard protocol, but the final shared state is not maximally entangled. When Alice and Bob want to obtain a closer state to $|\Psi_{00}\rangle$, the best they can do is to correct the state through the operations $\hat{Z}_1$ and $\hat{X}_4$.}
	\label{fig:circuitME}
\end{figure}

In quantum state discrimination an observer aims at identifying states generated by a source. The set of states and generation probabilities are known by the observer beforehand. This must design the optimal identification strategy according to a predefined figure of merit. The case of a set formed by orthogonal states is the simplest one since the measurement of an observable suffices to identify all states deterministically and conclusively. Nonorthogonal states cannot be identified in this form and, consequently, different techniques are required. In the particular case of ME we are interested in minimizing the probability of erroneous identification on average or, equivalently, to achieve a maximal probability to make correct assignments about input states from the measurement results~\cite{Holevo73,HelstromBook}. The explicit form of the positive operator-valued measure (POVM) that allow this class of discrimination among equally likely symmetric states is known~\cite{Ban97,Barnett01}. In the scenario described above, it consists of an inverse Fourier transform acting on a $D_A$-dimensional Hilbert space followed by a projective measurement on system 2 onto the computational basis of this space. Therefore, the implementation of optimal ME strategy is equivalent to carry out the standard ES protocol described in Sec.~\ref{sec:standardES}, as depicted in Fig.~\ref{fig:circuitME}. Using Eqs.~(\ref{FourierTransformD}), (\ref{nustates}), and (\ref{Xis}), this procedure leads the state of systems 1 and 4 to be transformed into 

\begin{align}
	|\Phi_{ms}\rangle_{14} &= \frac{1}{\sqrt{D_{A}}}\sum_{l=0}^{D_{A}-1}\left(\sum_{n=0}^{D_{A}-1}\omega^{ n(l-m)}\Gamma_{ns}\right)|\Psi_{ls}\rangle_{14}. \label{Phims}
\end{align}
which is a superposition of maximally entangled states and reduces to $|\Psi_{ms}\rangle$ when $|\phi\rangle_A$ and $|\varphi\rangle_B$ are maximally entangled states. 

\par To quantify the entanglement of bipartite pure states $|\Psi\rangle_{14}$ we make use of the linear entropy of one of the reduced density matrices,
\begin{align}
	\mathcal{E}(|\Psi\rangle_{14}) = \frac{\left[1 - {\rm tr}\left(\hat\rho_1^2\right)\right]D_A}{D_A-1} = \frac{\left[1 - {\rm tr}\left(\hat\rho_4^2\right)\right]D_A}{D_A-1},\label{ent_measure1} 
\end{align}
where $\hat\rho_1$ ($\hat\rho_4$) is the reduced density operator of system $1$ (system $4$). This function satisfies the necessary conditions for an entanglement measure~\cite{Vedral97} and is the square of the I-concurrence \cite{Rungta01}. This measure adopts values between 0 and 1. Hence, the entanglement of $|\Phi_{ms}\rangle_{14}$ characterized by the measure function of Eq.~(\ref{ent_measure1}) is 
\begin{align}
	\mathcal{E}_{ms}^{\rm ME} &= \frac{D_A}{D_A-1}\left(1-\sum_{q=0}^{D_A-1}\Gamma_{qs}^4\right) .\label{ent_MEms}
\end{align}

Additionally, after obtaining the outcomes $m$ and $s$ in the measurement of systems 2 and 3, respectively, Charlie could assume that the shared state is closer to the state $|\Psi_{ms}\rangle_{14}$ than to the other maximally entangled states of Eq.~(\ref{Psistates}). Accordingly, he could compute the fidelity between the state (\ref{Phims}) and the maximally entangled state $|\Psi_{ms}\rangle$ generated in a perfect ES protocol. Doing so he will obtain 
\begin{align}\nonumber
	\mathcal{F}_{ms}^{\rm ME} &=  \left| {}_{14}\langle \Psi_{ms}|\Phi_{ms}\rangle_{14}\right|^2\\
	&= \frac{1}{D_{A}}\left( \sum_{q=0}^{D_{A}-1}\Gamma_{qs} \right)^{2}.\label{ent_fidME}
\end{align}

\par These figures of merit were calculated considering any particular outcome of the measurements on particles 2 and 3. Note that their values depend on the index $s$ that determines the set of symmetric states to be discriminated  and not on the index $m$ that labels the states within a certain set. Indeed, these figures depend on the modulus of the complex coefficients of the states to be discriminated, which adopt the same values for every vector within a given set. Following the example of Fig.~\ref{fig:circuit1}, assume that Alice and Bob would like to share a $|\Psi_{00}\rangle$ state. Since $|\Phi_{ms}\rangle$ is not maximally entangled, the best they can do is to transform this state into a closer one to the desired state. From Eqs.~(\ref{Psistates}) and~(\ref{Phims}) we have that 
\begin{equation}
	{}_{14}\langle \Psi_{ms}|\Phi_{ms}\rangle_{14} = {}_{14}\langle \Psi_{00}|~\hat{Z}_1^m \hat{X}_4^s|\Phi_{ms}\rangle_{14},
\end{equation}
which can be interpreted as a preservation of the fidelity of Eq.~(\ref{ent_fidME}), respect to the state $|\Phi_{00}\rangle$, when Alice applies $\hat{Z}_1^m$ and Bob applies $\hat{X}_4^s$ on $|\Phi_{ms}\rangle$, as Fig.~\ref{fig:circuitME} shows. Since these operations are unitary and local, they do not modify the obtained entanglement.
\par An interesting quantity is the average of these figures of merit over every possible attainable result $(m,s)$, which gives us an overall figure of merit of ES assisted by ME discrimination. We define the average entanglement as the sum of the entanglement obtained at each combination of $s$ and $m$ weighted by the total probability of having obtained this pair of outcomes. Then, considering Eq.~(\ref{ent_MEms}), we have 
\begin{align} 
\langle\mathcal{E}^{\rm ME}\rangle 
	&= \frac{D_A}{D_A-1}\left(1 - \sum_{s=0}^{D_B-1} p_s \sum_{q=0}^{D_A-1} \Gamma_{qs}^4\right),
\label{eq:MEentav}\end{align}
where $p_s$ and $\Gamma_{qs}$ are given below Eqs.~(\ref{xorphiphi}) and~(\ref{nustates}), respectively. Analogously as it was done with the entanglement, the average for the fidelities of Eq.~(\ref{ent_fidME}) is given by 
\begin{align} 
\langle\mathcal{F}^{\rm ME}\rangle 
	&= \frac{1}{D_A}\sum_{s=0}^{D_B-1}p_s\left(\sum_{q=0}^{D_A-1} \Gamma_{qs}\right)^2.
\label{eq:MEfidav}\end{align}
Let us note that ES assisted by ME is a deterministic process since ME always succeed in spite of the fact that the final state is not maximally entangled.

\subsection{ES assisted by MC discrimination \label{sec:MC}}

Now, assume that Charlie, in agreement with Alice and Bob, will apply MC strategy for the discrimination of the states $|\nu_{ls}\rangle$ in Eq.~(\ref{nustates}).\footnote{Since the linear dependence of the states $|\nu_{ls}\rangle$ can vary for every set of states $\Omega_s$, the UD strategy cannot be employed here as a general rule. Nevertheless, MC reduces to UD when the states to be discriminated are linearly independent~\cite{BergouReview}. Moreover, it has been reported that probabilistic enhancements of other quantum protocols can be obtained when this class of measurements is applied as, for instance, quantum state teleportation via pure nonmaximally entangled states~\cite{Neves12, SolisProsser13}.} This is a probabilistic process that leads to conclusive events, which allows one to identify the state with the \emph{maximum achievable confidence}, and inconclusive events, in which such attempt fails~\cite{Croke06,Jimenez11,Herzog12}. Therefore, when MC measurements are applied to the ES protocol the parties are aware of the chance of failure. However, as we show in this section, where only conclusive (or successful) events in the discrimination process are considered, both entanglement and fidelity are larger than the ones achieved in the deterministic protocol described above. Besides, these figures of merit depend only on the number of nonvanishing coefficients of the states $|\nu_{ms}\rangle$ for every given result $(m,s)$.

The optimal POVM for discriminating among equally likely symmetric states via MC has been derived in Ref.~\cite{Jimenez11}. In order to implement it, Charlie must introduce a two-dimensional ancillary system (ancilla) initially prepared in the state $|0\rangle_a$. In the first step, conditional to the outcome $s$ in the measurement of system 3, he applies a unitary operation $\hat{U}_{2a}$ that couples system 2 and ancilla as 
\begin{align}
	\hat{U}_{2a}^{(s)}|\nu_{ls}\rangle_{2}|0\rangle_{a} &= \hat{A}_{\rm succ}^{(s)}|\nu_{ls}\rangle_{2}|0\rangle_{a} + \hat{A}_{\rm ?}^{(s)}|\nu_{ls}\rangle_{2}|1\rangle_{a},\label{eqPOVM1}
\end{align}
where the operators $\hat{A}_{\rm succ}^{(s)}$ and $\hat{A}_{\rm ?}^{(s)}$ are defined as
\begin{align}
	\hat{A}_{\rm succ}^{(s)} =& \sum_{q=0}^{D_{A}-1}\frac{\gamma_{s}}{\Gamma_{qs}}y_{qs}|q\rangle_{2}\langle q|,\\
	\hat{A}_{\rm ?}^{(s)} =& \sum_{q=0}^{D_{A}-1}\sqrt{1 - \frac{\gamma_{s}^{2}}{\Gamma_{qs}^{2}} }y_{qs}|q\rangle_{2}\langle q|,
\end{align}
with ${\gamma_{s} = \min_{}\{\Gamma_{js}: 0\leqslant j \leqslant D_A-1 ~~{\rm and}~~\Gamma_{js}\ne 0\},}$ and ${\hat{A}_{\rm succ}^{(s)\dagger}\hat{A}_{\rm succ}^{(s)} + \hat{A}_{\rm ?}^{(s)\dagger}\hat{A}_{\rm ?}^{(s)}=\hat{1}_{D_A}}$. We have defined $y_{qs}$ and $1/\Gamma_{qs}$ such that 
\begin{align}
	y_{qs} &= \begin{cases}1 & {\rm if~~~} \Gamma_{qs}\ne 0\\ 0 & {\rm if~~~} \Gamma_{qs}= 0 \end{cases}, \\
	\frac{y_{qs}}{\Gamma_{qs}} &=\begin{cases}\frac{1}{\Gamma_{qs}} & {\rm if~~~} \Gamma_{qs}\ne 0\\ 0 & {\rm if~~~} \Gamma_{qs}= 0 \end{cases}.
\end{align}

\begin{figure}[!t]
	\centering
	\includegraphics[width=0.45\textwidth]{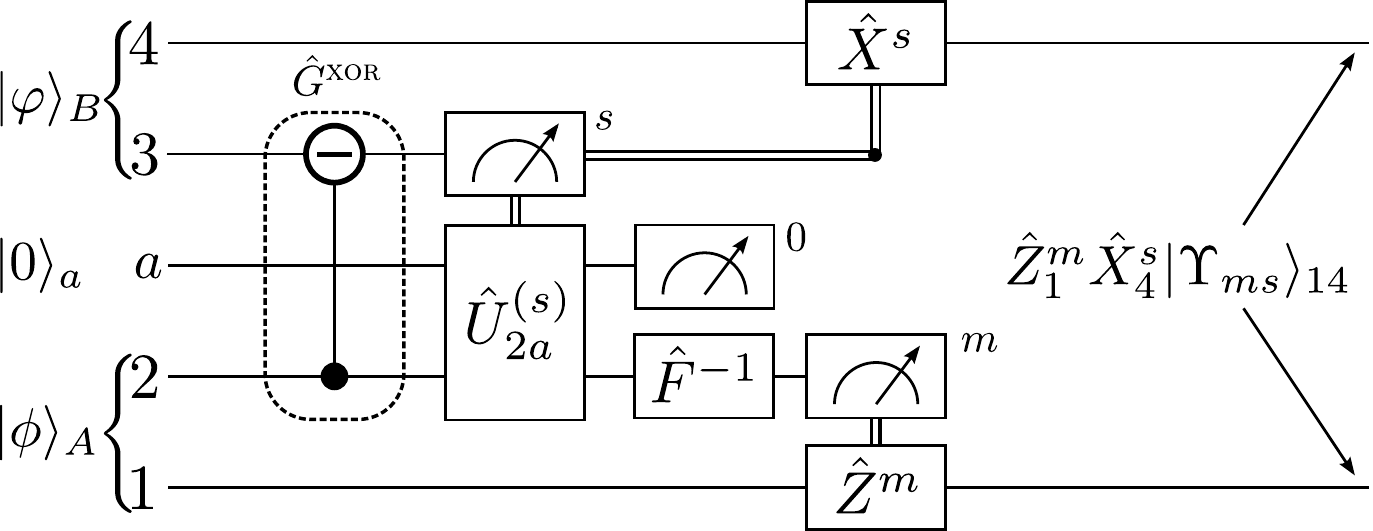}
	\caption{Example of ES assisted by MC measurements. Alice and Bob can apply unitary transformations $\hat{Z}_1$ and $\hat{X}_4$ in order to obtain a state closer to $|\Psi_{00}\rangle$. This diagram considers a successful MC discrimination as the ancilla was found in the state $|0\rangle$. \label{fig:circuitMCsuc}}
\end{figure}

As a consequence of Eq.~(\ref{eqPOVM1}), assisting the ES protocol via MC is equivalent to performing a probabilistic map on the tripartite states~(\ref{Xis}) whose result is determined by the state in which the ancilla is found. In general terms, this map is given by 
\begin{align}
	\hat{U}_{2a}^{(s)}|\Xi_{s}\rangle_{124}|0\rangle_{a} =& \sqrt{p_{\rm succ}^{(s)}/D_A}\sum_{l=0}^{D_{A}-1}|\Psi_{ls}\rangle_{14}|u_{ls}\rangle_{2}|0\rangle_a \nonumber\\ & + \sqrt{p_{\rm ?}^{(s)}/D_A}\sum_{l=0}^{D_{A}-1}|\Psi_{ls}\rangle_{14}|\nu_{ls}^{\prime}\rangle_{2}|1\rangle_a, \label{after_u2a}
\end{align}
where the states $|u_{ls}\rangle_{2}$ and $|\nu_{ls}^{\prime}\rangle_{2}$ are defined as 
\begin{align}
	|u_{ls}\rangle_{2} &= \frac{1}{\sqrt{\mathcal{N}_{s}}}\sum_{q=0}^{D_{A}-1} y_{qs} \omega^{ql} |q\rangle_{2} , \label{ustates}\\
	|\nu_{ls}^{\prime}\rangle_{2}  &= \frac{1}{\sqrt{p_{\rm ?}^{(s)}}}\sum_{q=0}^{D_{A}-1} y_{qs} \omega^{ql} \sqrt{ \Gamma_{qs}^{2} - \gamma_{s}^{2}  }|q\rangle_{2}, \label{nu1states}
\end{align}
respectively, and 
\begin{equation}
	p_{\rm succ}^{(s)}=\mathcal{N}_s\gamma_s^2 = 1-p_{\rm ?}^{(s)},
\end{equation}
where $\mathcal{N}_{s} = \sum_{q=0}^{D_{A}-1}y_{qs}$ is the number of nonvanishing coefficients $\Gamma_{qs}$. The quantity $p_{\rm succ}^{(s)}$ $(p_{\rm ?}^{(s)}$) represents the maximum (minimum) probability of obtaining a conclusive (inconclusive) result. The first step of Charlie's measurement on system 2 is accomplished by measuring the ancilla on the computational basis. When it is found in $|0\rangle_a$, the measurement is considered successful (or conclusive). In this case, the second and final step of the MC measurement comprises a ME measurement~\cite{Jimenez11}, i.e., Charlie applies an inverse Fourier transform on the state $|u_{ls}\rangle$ followed by a projective measurement onto the computational basis, both acting on a $D_A$-dimensional Hilbert space. Using Eqs.~(\ref{FourierTransformD}), (\ref{after_u2a}), and~(\ref{ustates}), and assuming that he obtained the $m$th outcome when he measured system 2, this procedure leads the state of systems 1 and 4 to be transformed into  

\begin{align}
	|\Upsilon_{ms}\rangle_{14}=\frac{1}{\sqrt{\mathcal{N}_{s}D_A}} \sum_{l=0}^{D_{A}-1}\left[ \sum_{n=0}^{D_{A}-1}y_{ns} \omega^{ n(l-m)}\right]|\Psi_{ls}\rangle_{14}.\label{upsilon}
\end{align}
Figure~\ref{fig:circuitMCsuc} shows a simplified quantum circuit explaining the process that integrates MC measurements for assisting the ES protocol. The entanglement of this state is, according to Eq.~(\ref{ent_measure1}), 
\begin{align}
	\mathcal{E}_{ms}^{\rm MC} &= \frac{D_A}{D_A-1}\left(1 - \frac{1}{\mathcal{N}_{s}}\right).\label{ent_MCIms}
\end{align}
On the other hand, the fidelity of this state respect to the ideal maximally entangled one generated in a perfect ES protocol [see Eq.~(\ref{Psistates})] is
\begin{align}
	\mathcal{F}_{ms}^{\rm MC} &= |{}_{14}\langle\Psi_{ms}|\Upsilon_{ms}\rangle_{14}|^{2} = \frac{\mathcal{N}_{s}}{D_{A}}.\label{ent_fidMC}
\end{align}

Thus, when the number of non-zero coefficients of $|\nu_{ls}\rangle_2$ is equal to $D_A$, we obtain $\mathcal{E}_{ms}^{\rm MC}=1$ and $\mathcal{F}_{ms}^{\rm MC}=1$, which are the maximum values attainable for these quantities, reproducing the results found in Ref.~\cite{Delgado05}.

It is noteworthy that the results of Eqs.~(\ref{ent_MCIms}) and~(\ref{ent_fidMC}) depend on the number of nonvanishing coefficients $\Gamma_{ms}$ of the states $|\nu_{ls}\rangle$ of a given set $s$ resulting from a measurement on system 3. There is no dependence on their numerical values themselves as in the case of a deterministic protocol [see Eqs.~(\ref{ent_MEms}) and~(\ref{ent_fidMC})]. This is a nice feature of MC measurements among equally likely symmetric states, where our confidence for identifying each state in the set depends only on the dimension of the subspace spanned by the states ($\mathcal{N}_s$) and the number of states in the set ($D_A$), through the ratio $\mathcal{N}_s/D_A$   \cite{Jimenez11,Herzog12}.

\begin{figure*}[!t]
	\centering
	\includegraphics[width=0.95\textwidth]{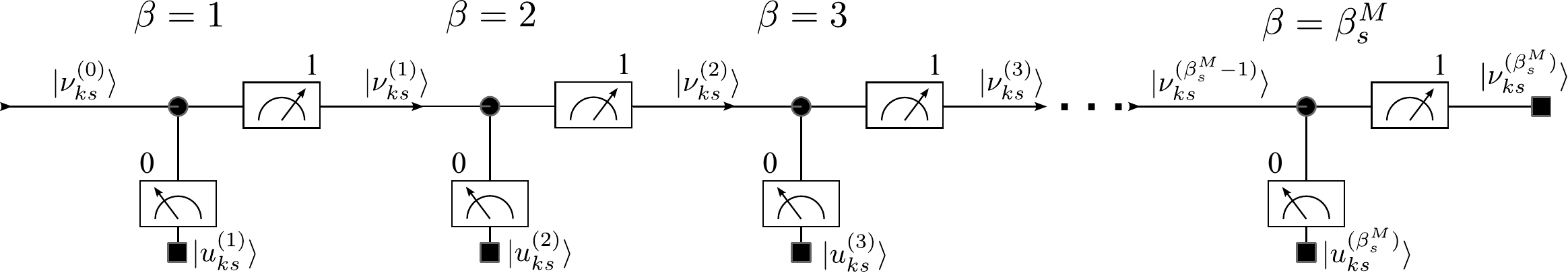}
	\caption{A simplified scheme of SMC measurements. The system carrying the state $|\nu_{ls}^{(0)}\rangle$ to be discriminated is made interact with a two-dimensional ancilla. Afterwards, the ancilla is measured onto the computational basis. When the measurement fails ($``1"$), states $|\nu_{ks}^{(0)}\rangle$ are mapped on $|\nu_{ks}^{(1)}\rangle$ and additional stages can be implemented attempting to extract the most confident remaining information until the maximum number of MC stages is reached. If the process continues failing, the states are mapped onto $|\nu_{ks}^{(\beta)}\rangle$ according the number $\beta$ of employed MC stages increases. When the process succeeds ($``0"$) at any given $\beta$th stage, the states $|\nu_{ks}^{(\beta-1)}\rangle$ are mapped on $|u_{ks}^{(\beta)}\rangle$, which will be discriminated through ME. \label{SMC}}
\end{figure*}

\subsubsection*{Comparison between ME and MC for the figures of merit $\mathcal{E}$ and $\mathcal{F}$\label{comparison1}}
\par First, we can compare the fidelity between the actually obtained entangled state and the ideal maximally entangled state from a perfect ES protocol for both discrimination strategies. Although this is not properly an entanglement measure, it is interesting to analyze how close Alice and Bob are to sharing such ideal state. From Eq.~(\ref{ent_fidME}) we have that 
\begin{align}
	\mathcal{F}_{ms}^{\rm ME} &= \frac{1}{D_A}\left|\sum_{q}\Gamma_{qs}\right|^2 = \frac{1}{D_A}\||\nu_{0s}\rangle\|_1^2,
\end{align}
where the symbol $\|\cdot\|_1$ denotes the $\ell^1$-norm of a vector. It can be shown, aided by the Schwarz inequality, that the $\ell^1$-norm and the usual $\ell^2$-norm induced by the canonical Hermitian product, which we shall denote by $\|\cdot\|_2$, must satisfy ${\|\mathbf{u}\|_1\leqslant \sqrt{n}\|\mathbf{u}\|_2}$ for every vector $\mathbf{u}$ belonging to the Hilbert space, where $n$ is the number of nonvanishing coefficients of $\mathbf{u}$. For the states $|\nu_{0s}\rangle$, $n$ will coincide with $\mathcal{N}_s$. This inequality, according to Eq.~(\ref{ent_fidMC}), implies that
\begin{align}
	\mathcal{F}_{ms}^{\rm ME} 
			& \leqslant \frac{1}{D_A}\left(\sqrt{\mathcal{N}}_s\||\nu_{0s}\rangle\|_2\right)^2 \nonumber\\
			&= \frac{\mathcal{N}_s}{D_A} = 	\mathcal{F}_{ms}^{\rm MC}.
\end{align}

Now, we are going to compare the entanglement between these two techniques. Let us define an auxiliary vector $\mathbf{g}$ such that
\begin{align}
	\mathbf{g}_s &= \sum_{q=0}^{D_A-1}|\Gamma_{qs}|^2\mathbf{e}_q,
\end{align}
This definition implies that ${\|\mathbf{g}\|_1^2=1}$ and $\|\mathbf{g}\|_2^2=\sum_{q=0}^{D_A}|\Gamma_{rs}|^4$. From Eq.~(\ref{ent_MEms}), we have that
\begin{align}
	\mathcal{E}_{ms}^{\rm ME} & =  \frac{D_A}{D_A-1}\left(1- \sum_{q=0}^ {D_A-1} |\Gamma_{qs}|^4\right) \nonumber\\
	& = \frac{D_A}{D_A-1}\left(1 - \|\mathbf{g}_s \|_2^2 \right).
\end{align}
Since $\|\mathbf{g}_s\|_2 \geqslant \| \mathbf{g}_s\|_1/\sqrt{\mathcal{N}_s} = 1/\sqrt{\mathcal{N}_s}$ and taking Eq.~(\ref{ent_MCIms}) into account, it can be proven that 
\begin{align}
		\mathcal{E}_{ms}^{\rm ME} & \leqslant \mathcal{E}_{ms}^{\rm MC}.
\end{align}

The above results show that a conclusive event in ES assisted by MC leads us to obtain bipartite states with equal
or higher entanglement and fidelity than the ones attainable through the standard procedure---in which ES is
assisted by ME discrimination---for each possible pair of outcomes $(m,s)$. Nevertheless, this advantage is attained
at the expense of moving from a deterministic process to a probabilistic one.

\subsection{Multistage MC assisted ES \label{sec:SMC}}

Unlike UD, quantum state discrimination via MC can be continued after a failed discrimination attempt~\cite{Jimenez11}. In this way, we can analyze the potential improvements of the ES protocol under these circumstances. For this purpose, let us consider that after applying the unitary operation (\ref{eqPOVM1}), Charlie's measurement on the ancilla projects it onto $|1\rangle_{a}$. In this case, the state of system 2 is mapped on the states $|\nu_{ls}^{\prime}\rangle_{2}$ given by Eq.~(\ref{nu1states}). These states form a new set of equally likely symmetric states  which, however, span a smaller subspace than the one spanned by the initial $|\nu_{ms}\rangle$ since the former have, at least, one additional null coefficient. Thus, they are amenable to a new stage of MC measurement as described above. This can be useful when Alice, Bob, and Charlie have limited resources (entangled states) and cannot afford simply discarding the particles after a failed attempt. 
\par The procedure when Charlie fails in the first MC discrimination attempt is to prepare again an ancilla in $|0\rangle_a$ to be coupled with system 2 afterward, followed by a measurement on the ancilla that will determine whether this new attempt was conclusive. In the conclusive case, an inverse Fourier transform and a standard projective measurement on the computational basis are performed on system 2. In case of a new failure, the obtained states after two failed attempts can be either discriminated with ME or, when possible, with an additional MC procedure (i.e. by coupling with an ancilla and subsequent measurement on it). This gives rise to the \emph{sequential maximum-confidence} (SMC) measurements, as shown in Fig.~\ref{SMC}. 
For this strategy, we shall define some helpful quantities to describe the coefficients and probabilities of the probabilistically transformed states. First, an additional label $\beta$ will be introduced to indicate how many MC stages have been passed through. Then, for $\beta\geqslant 1$, we define the parameters

\begin{subequations}
\begin{align}
	\Gamma_{qs}^{(\beta)} &= \left[\frac{\left(\Gamma_{qs}^{(\beta-1)}\right)^2 - \left(\gamma_{s}^{(\beta-1)}\right)^2}{1 - \mathcal{N}_s^{(\beta-1)}\left(\gamma_{s}^{(\beta-1)}\right)^2 }\right]^{1/2}, \\
	\gamma_s^{(\beta-1)}&=\min\left\{\Gamma_{qs}^{(\beta-1)}~: ~\Gamma_{qs}^{(\beta-1)}\neq 0\right\} ,\\
	|\nu_{ls}^{(\beta)}\rangle &= \sum_{q=0}^{D_A-1} y_{qs}^{(\beta-1)}\Gamma_{qs}^{(\beta)}\omega^{ql}|q\rangle,\\
	|u_{ls}^{(\beta)}\rangle &= \tfrac{1}{\sqrt{\mathcal{N}_s^{(\beta-1)}}}\sum_{q=0}^{D_A-1} y_{qs}^{(\beta-1)}\omega^{ql}|q\rangle ,\\
	y_{qs}^{(\beta-1)}& =\begin{cases} 1 ~~~;~~~	\Gamma_{qs}^{(\beta-1)}\neq 0 \\ 0 ~~~;~~~	\Gamma_{qs}^{(\beta-1)} = 0\end{cases}, \\
	\mathcal{N}_s^{(\beta-1)}&=\sum_{q=0}^{D_A-1}y_{qs}^{(\beta-1)}.
\end{align}
\end{subequations}

The interpretation of these parameters is as follows: $\Gamma_{qs}^{(\beta)}$ are the coefficients of the states $|\nu_{ls}^{(\beta)}\rangle$ to be discriminated after $\beta$ failed MC attempts and $\gamma_{s}^{(\beta)}$ is the smallest among them. The states $|u_{ls}^{(\beta)}\rangle$ correspond to the states obtained after ${\beta-1}$ failed MC attempts and a last successful MC operation. The binary parameter $y_{qs}^{(\beta-1)}$ contains information about the non-nullity of each $\Gamma_{qs}^{(\beta-1)}$ and indicates which coefficients of the states $|\nu_{ls}^{(\beta)}\rangle$ will not vanish. Finally, $\mathcal{N}_s^{(\beta-1)}$ is the total number of nonvanishing coefficients of these states.\footnote{%
Furthermore, in terms of previously defined quantities, we can regard %
$|\nu_{ls}\rangle_2 $, $y_{qs} $, $\Gamma_{qs}$, and $ \mathcal{N}_s$ as corresponding for $\beta=0$, meanwhile $|\nu_{ls}^{\prime}\rangle_2 $ and $|u_{ls}\rangle_2 $ are for $\beta=1$} %
The implementation of each $\beta$th stage requires an ancilla prepared in $|0\rangle_a$ and a new bipartite unitary operation such that 
\begin{align}
	\hat{U}_{2a}^{(s,\beta)}|\nu_{ls}^{(\beta-1)}\rangle_{2}|0\rangle_{a} =& \sqrt{p_{\rm succ}^{(s,\beta)}}|u_{ls}^{(\beta)}\rangle_{2}|0\rangle_{a} \nonumber\\ &+ \sqrt{1-p_{\rm succ}^{(s,\beta)}}|\nu_{ls}^{(\beta)}\rangle_{2}|1\rangle_{a},\label{u2beta}
\end{align}
where 
\begin{align}
	p_{\rm succ}^{(s,\beta)} = \mathcal{N}_{s}^{(\beta-1)}\left[\gamma_{s}^{(\beta-1)}\right]^2,
\end{align}
represents the probability of having a conclusive event in the $\beta$th stage. Since the measurements are performed only on the ancilla, the original system $2$ and its quantum state will not be destroyed along this process until a final ME discrimination measurement is carried out at any chosen stage. 

\begin{figure*}[t!]
	\centering
	\includegraphics[width=0.95\textwidth]{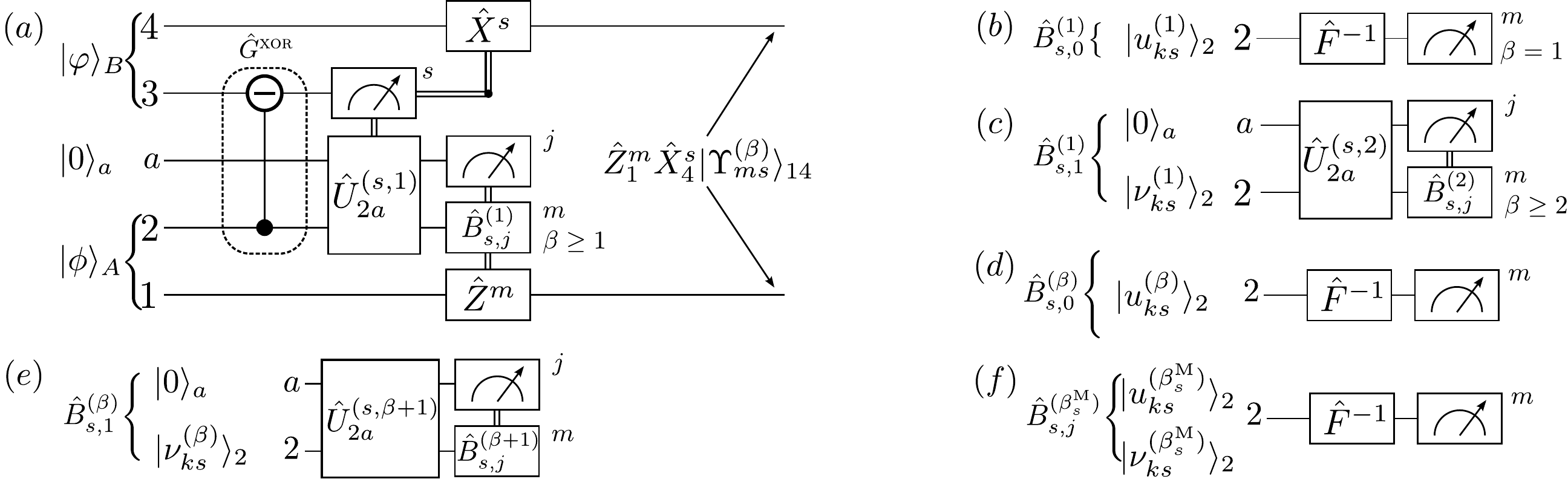}
	\caption{ES circuit assisted by sequential maximum-confidence (SMC) measurements. (a) Charlie attempts to discriminate among the set $\{|\nu_{ms}\rangle\}$ after having measured on particle 3. The bipartite unitary operation $\hat{U}_{2a}^{(s,1)}$ and the measurement on the ancilla yield two options, $j=0,1$. Unitary operations made by Alice and Bob allow us to bring the final shared state closer to $|\Psi_{00}\rangle$ if required. (b) If the process were successful, when $j=0$, the final step of the discrimination is performed and only the $\beta=1$ stage of MC measurements was needed. (c) When $j=1$, Charlie can try again to discriminate, although less confidently, by an additional MC stage. (d) For any $\beta$th stage, in case of success, Charlie can apply ME discrimination among the states $|u_{ks}^{(\beta)}\rangle$ and report the result. (e) In case of failure at the $\beta$th stage, it is possible to try to discriminate again via MC, having used $\beta+1$ stages at least. (f) When the number of employed stages reaches the maximum chosen value $\beta_{s}^{\rm M}$, despite having success ($j=0$) or not ($j=1$), ME can be applied. \label{fig:circuitSMC}}
\end{figure*}

\begin{figure*}[tb]
	\includegraphics[width=0.95\textwidth]{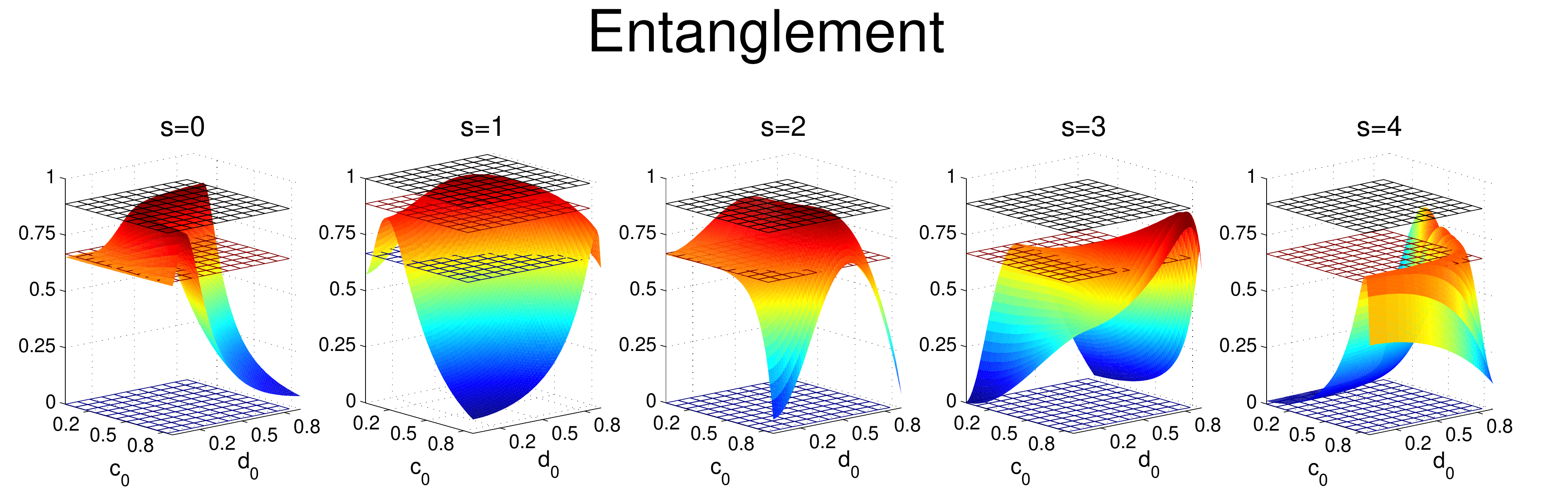}\vspace{0.5cm}
	\includegraphics[width=0.95\textwidth]{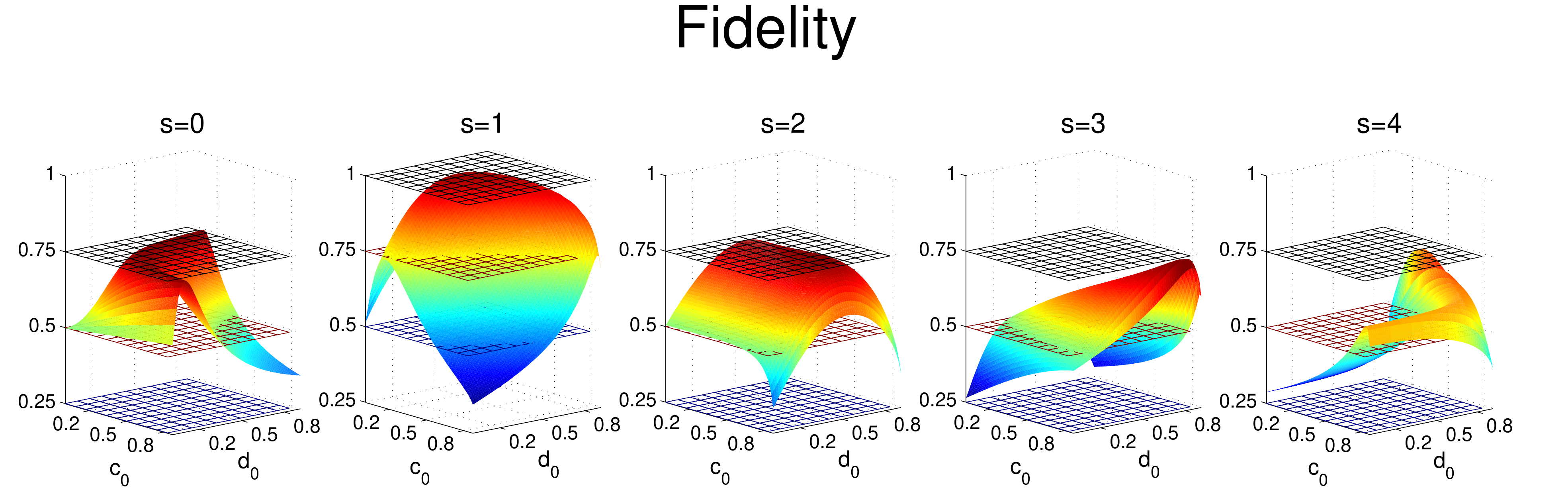}
	\caption{(Color online) A simulation of the attainable entanglement (upper panels) and fidelity (lower panels) via ME (curved surface) and MC (horizontal planes) for $D_A=4$ and $D_B=5$. The parameters $c_0$ and $d_0$ are free. We have set $c_1 = 0.2811$, $c_2 = 0.3790$, $d_1 = 0.3220$, $d_2 = 0.2064$ and $d_3 = 0$. Coefficients $c_3$ and $d_4$ are obtained by normalization. 
     \label{fig:entandfid}}
\end{figure*}

In addition to the parameters defined above, we must bear in mind the multiplicity of the coefficients in the states to be discriminated. For this purpose, let us consider---only for now---a rearrangement of the coefficients of the fiducial states $|\nu_{0s}\rangle$ such that they look increasingly ordered. Since we will have $n_s\leqslant D_A$ different values for the coefficients (degeneracy decreases this number), it will be necessary to define $\mu_{ws}$ as the multiplicity of the $w$-th smallest value for the state $|\nu_{0s}\rangle$, where $w=1,\dots,n_s$. For instance, $\mu_{1s}$ will be the multiplicity of the smallest coefficient, $\mu_{2s}$ will correspond to the second-smallest coefficient multiplicity, and $\mu_{n_s s}$ will indicate how many times the largest coefficient is repeated. Clearly, $\sum_{w=1}^{n_s}\mu_{ws}=\mathcal{N}_s$ and $\sum_{w=1+\beta}^{n_s}\mu_{ws}=\mathcal{N}_s^{(\beta)}$. Recapitulating, as the circuits of Fig.~\ref{fig:circuitSMC} show, Charlie has measured system $3$ and found it in the state $|s\rangle_3$. He knows that the state of system 2 is one among the states $|\nu_{ls}\rangle$, with $s$ determined by his previous measurement. According to his goals, he can now perform ME discrimination, which coincides with the standard protocol, or perform SMC measurements where the number of stages will be determined by a certain minimum value required for entanglement or fidelity. As we have previously shown, implementing MC starting from the states $|\nu_{ls}\rangle$ enables, in the successful case, to map these states onto the $|u_{ls}^{(1)}\rangle$ ones [Fig.~\hyperref[fig:circuitSMC]{\ref*{fig:circuitSMC}(b)}] and to swap with entanglement~$\propto(1-1/\mathcal{N}_s)$ and fidelity~$\mathcal{N}_s/D_A$. In case of failure, the states $|\nu_{ls}\rangle$ will be mapped to $|\nu_{ls}^{(1)}\rangle$. Since these states are also symmetric and equally likely, analogous physical deductions hold. Then, a second MC measurement [Fig.~\hyperref[fig:circuitSMC]{\ref*{fig:circuitSMC}(c)}] yields, when successful, entanglement~$\propto(1-1/\mathcal{N}_s^{(1)})$ and fidelity~$\mathcal{N}_s^{(1)}/D_A$ after having applied $\hat{F}^{-1}$ and a standard measurement. Otherwise, in case of failure, Charlie obtains the states $|\nu_{ls}^{(1)}\rangle$ mapped onto $|\nu_{ls}^{(2)}\rangle$. This process can be repeated $n_s-1$ times at most or even fewer times if Charlie chooses to apply ME in an intermediate stage and stop performing MC or simply stopping discriminating at all. So, when the $\beta$th MC measurement is successful, the obtained entanglement and fidelity will be\footnote{Let us note that, according to the notation used in Eq.~(\ref{EntFidBeta}), we can associate ${\mathcal{E}_{ms}^{\rm MC}=\mathcal{E}_{ms}^{\rm (1)}}$, ~${\mathcal{E}_{ms}^{\rm ME}= \mathcal{E}_{ms}^{\rm (0)}}$, ~${\mathcal{F}_{ms}^{\rm MC}=\mathcal{F}_{ms}^{\rm (1)}}$, and ~${\mathcal{F}_{ms}^{\rm ME}=\mathcal{F}_{ms}^{\rm (0)}}$. }
\begin{align}
	\mathcal{E}_{ms}^{(\beta)}= \frac{D_A}{D_A-1}\left(1 - \frac{1}{ \mathcal{N}_s^{(\beta-1)}}\right),\hspace{1.0cm}\mathcal{F}_{ms}^{(\beta)}=\frac{\mathcal{N}_s^{(\beta-1)}}{D_A}.\label{EntFidBeta}
\end{align}
Since $\sum_{w=1+\beta}^{n_s}\mu_{ws}=\mathcal{N}_s^{(\beta)}$, these quantities can be written in terms of the multiplicities defined for the coefficients of the fiducial states. So, the entanglement can be written as
\begin{align}
	\mathcal{E}_{ms}^{(\beta)}&=\frac{D_A}{D_A-1}\left(1 - \frac{1}{ \sum\limits_{w=\beta}^{n_s}\mu_{ws}}\right),\label{SMCfigs1}
\end{align}
and the fidelity as
\begin{align}
	\mathcal{F}_{ms}^{(\beta)}&=\frac{1}{D_A}\sum_{w=\beta}^{n_s}\mu_{ws}, \label{SMCfigs2}
\end{align} 
respectively. In addition, the SMC measurements can increase the probability of succeeding in carrying out the ES protocol outperforming the standard procedure. If $\beta_s^{\rm M}$ is the maximum number of SMC stages for a given value of $s$, the total probability of having had success before reaching the $(\beta_s^{\rm M}+1)$-th stage is
\begin{align}
	p_{{\rm succ}}^{(s,~\leqslant \beta_s^{\rm M})} = \sum_{\xi=1}^{\beta_s^{\rm M}} p_{\rm succ}^{(s,\xi)}\prod_{\lambda=1}^{\xi-1}\left(1-p_{\rm succ}^{(s,\lambda)}\right).\label{prob_succbeta}
\end{align}

\par Then, Charlie can establish a threshold for either entanglement or fidelity to be achieved as a minimum. Equations~(\ref{SMCfigs1}) and~(\ref{SMCfigs2}) allow him to decide how many SMC stages it is worth to implement according to the previously obtained result for $|s\rangle_3$ and decide a maximum number $\beta_{s}^{\rm M}$ of stages. This process has a success probability given by Eq.~(\ref{prob_succbeta}). Finally, ME can be implemented when any of the stages yield a successful result [Fig.~\hyperref[fig:circuitSMC]{\ref*{fig:circuitSMC}(d)}] or, in case of a failed operation and depending on experimental resources, Charlie can choose whether continue attempting through MC measurements [Fig.~\hyperref[fig:circuitSMC]{\ref*{fig:circuitSMC}(e)}] or to implement ME at the last allowed MC stage [Fig.~\hyperref[fig:circuitSMC]{\ref*{fig:circuitSMC}(f)}], or simply to discard the current systems and reattempt the process with a new couple of entangled pairs. Unless the latter option were chosen, Alice and Bob can apply unitary operations on their particles in a pursuit of having a state closer to the state $|\Psi_{00}\rangle_{14}$, as Fig.~\hyperref[fig:circuitSMC]{\ref*{fig:circuitSMC}(a)} shows, after a pair $(m,s)$ of outcomes is obtained and communicated.
\par Additionally, as was done with the ME strategy, the entanglement and fidelity can be averaged over every attainable result of the several measurements along the sequential process. Then, as a function of the maximum number $ \beta_{s}^{\rm M}$ of stages implemented, the average entanglement and fidelity of the shared states is 
\begin{widetext}
	\begin{align}
		\left\langle \mathcal{E}_{\rm MC}^{\left(\beta_{s}^{\rm M}\right)}\right\rangle =& \sum_{s=0}^{D_B-1}p_s \left\{ p_{\rm succ}^{(s,1)}\mathcal{E}_{0s}^{(1)} + \sum_{\beta=2}^{\beta_{s}^{\rm M}}\left[ \prod_{\xi=1}^{\beta-1}\left(1 - p_{\rm succ}^{(s,\xi)}\right)\right]p_{\rm succ}^{(s,\beta)}\mathcal{E}_{0s}^{(\beta)} + \left[ \prod_{\beta=1}^{\beta_{s}^{\rm M}}\left(1 - p_{\rm succ}^{(s,\beta)}\right)\right]\frac{\left[1-\displaystyle\sum_{q=0}^{D_A-1} \left(\Gamma_{qs}^{\left(\beta_{s}^{\rm M}\right)}\right)^4\right] }{(D_A -1)/D_A} \right\}, \label{eq:MCentav}\\
		\left\langle \mathcal{F}_{\rm MC}^{\left(\beta_{s}^{\rm M}\right)}\right\rangle =& \sum_{s=0}^{D_B-1}p_s \left\{ p_{\rm succ}^{(s,1)}\mathcal{F}_{0s}^{(1)} + \sum_{\beta=2}^{\beta_{s}^{\rm M}}\left[ \prod_{\xi=1}^{\beta-1}\left(1 - p_{\rm succ}^{(s,\xi)}\right)\right]p_{\rm succ}^{(s,\beta)}\mathcal{F}_{0s}^{(\beta)} + \left[ \prod_{\beta=1}^{\beta_{s}^{\rm M}}\left(1 - p_{\rm succ}^{(s,\beta)}\right)\right]\frac{\left[\displaystyle\sum_{q=0}^{D_A-1} \Gamma_{qs}^{\left(\beta_{s}^{\rm M}\right)}\right]^2}{D_A} \right\},\label{eq:MCfidav}
	\end{align}
\end{widetext}
where we have included the last failed attempt at the $\beta_{s}^{\rm M}$th stage. When this occurs, the states $|\nu_{ls}\rangle$ are mapped on the $|\nu_{ls}{}^{\left(\beta_{s}^{\rm M}\right)}\rangle$ ones, and the associated figures of merit are analogous to the ones written in Eqs. (\ref{ent_MEms}) and (\ref{ent_fidME}) when ME is applied among them.

\section{Numerical results \label{numerical}}

\begin{figure}[t!]	\includegraphics[width=0.49\textwidth]{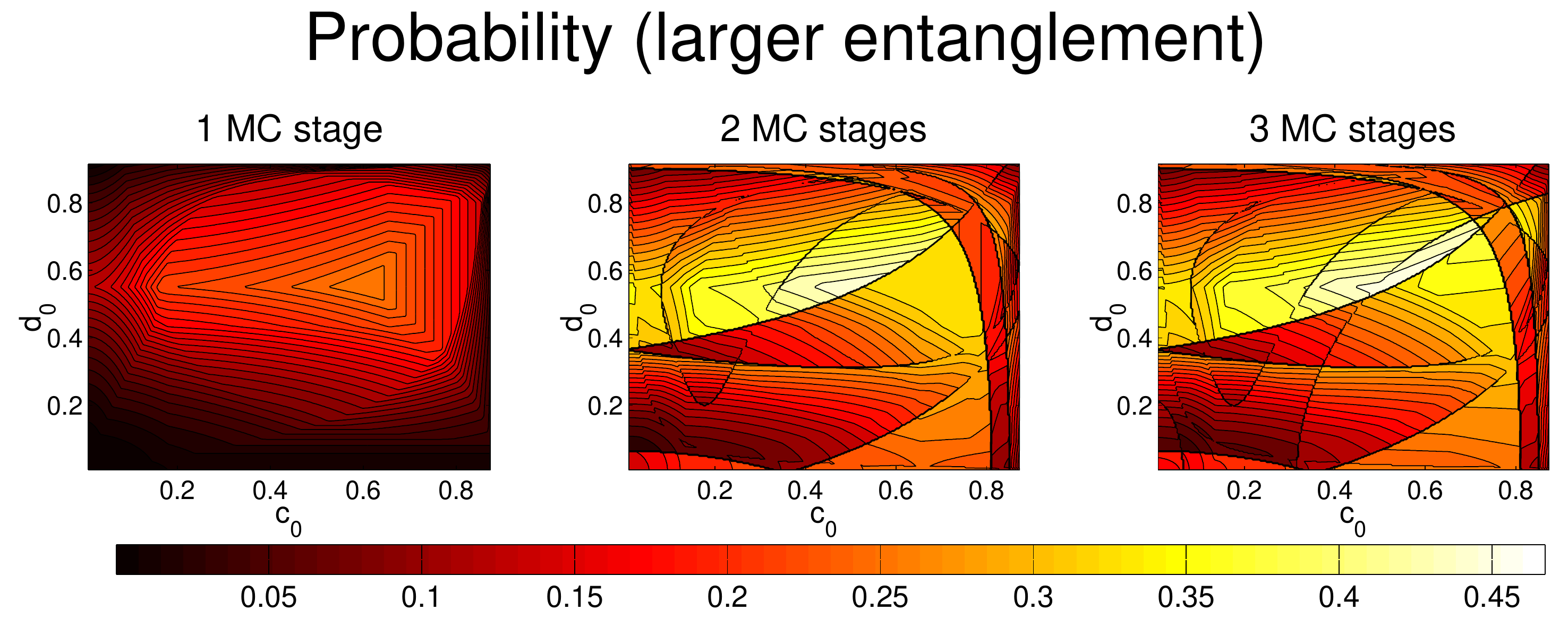}\vspace{0.5cm}
	\includegraphics[width=0.49\textwidth]{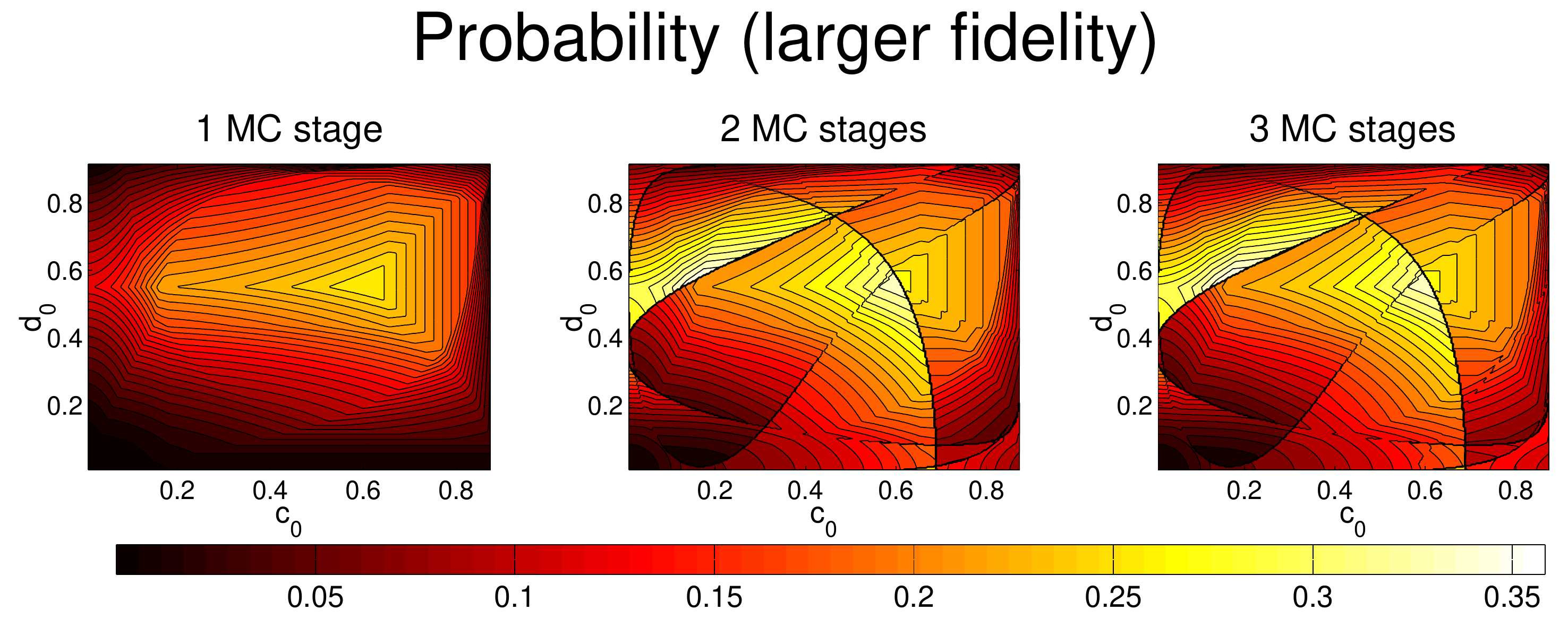}
	\caption{(Color online) A simulation of the total probability [Eq.~(\ref{prob_succbeta})] of swapping with higher entanglement (upper panels) and higher fidelity (lower panels) than the ones attainable by the standard protocol when the maximum number of allowed MC stages is increased. The same Schmidt coefficients pointed out in Fig.~\ref{fig:entandfid} were used. 
     \label{fig:Contour_probability}}
\end{figure}

\begin{figure}[t!]
	\includegraphics[width=0.49\textwidth]{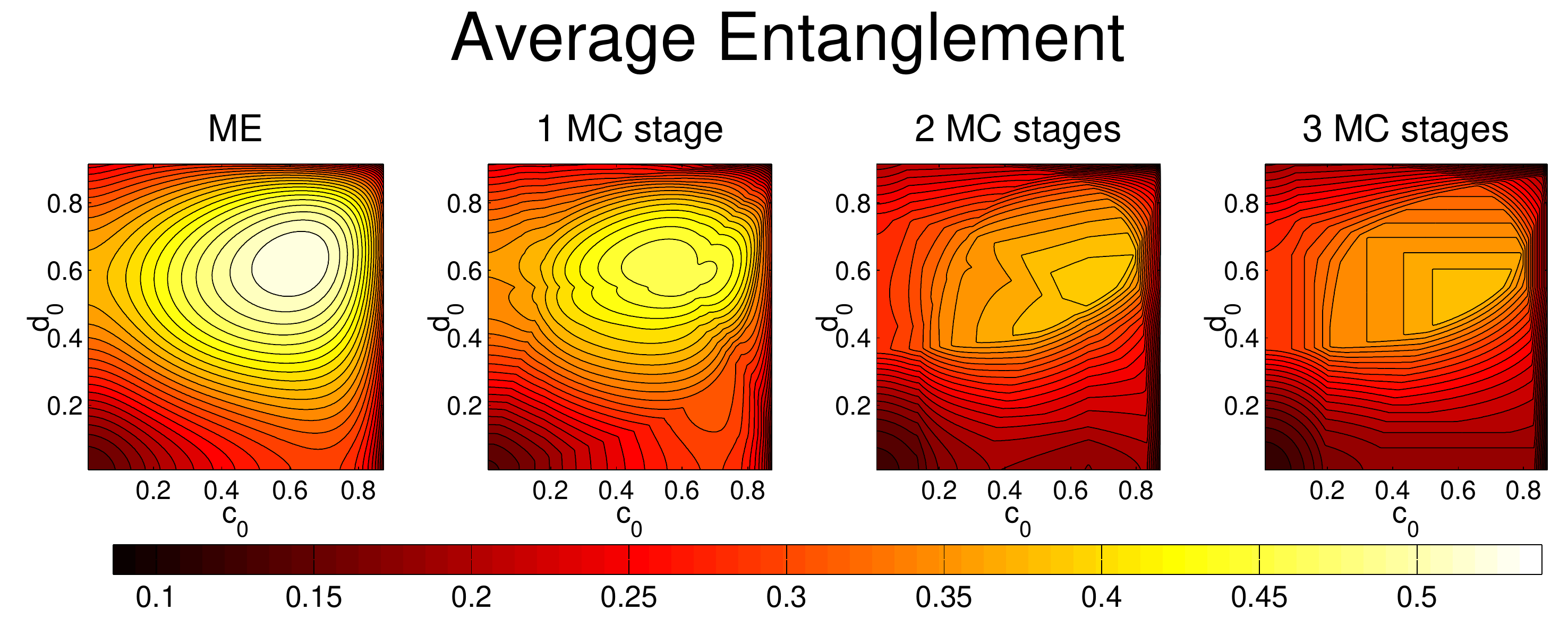}\vspace{0.5cm}
	\includegraphics[width=0.49\textwidth]{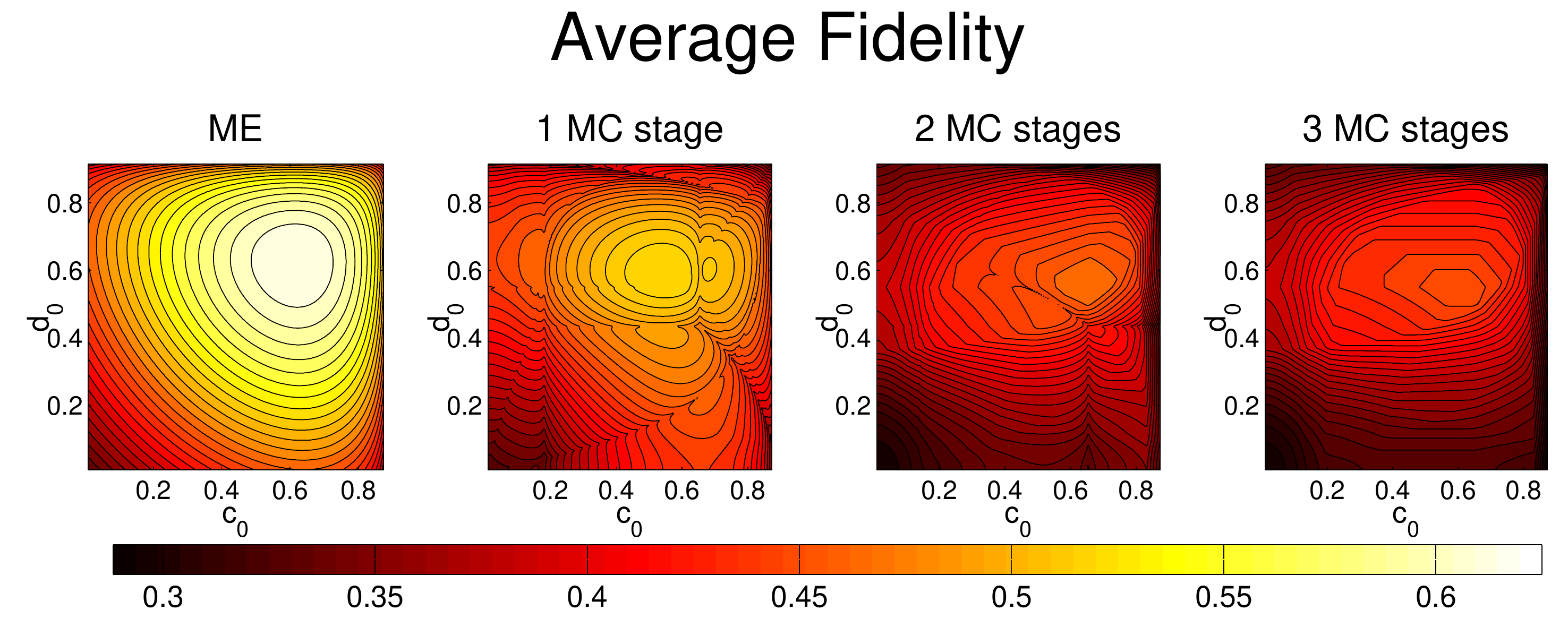}
	\caption{(Color online) A simulation of the overall average entanglement (upper panels) and fidelity (lower panels), considering every possible outcomes combination, when the number of MC stages implemented increased, i.e., $\beta_{s}^{\rm M}= 1, 2$, and $3$, for every $s$. The Schmidt coefficients are the same used for Figs.~\ref{fig:entandfid} and~\ref{fig:Contour_probability}. 
     \label{fig:Contour_average}}
\end{figure}

\begin{figure}[t!]
	\includegraphics[width=0.49\textwidth]{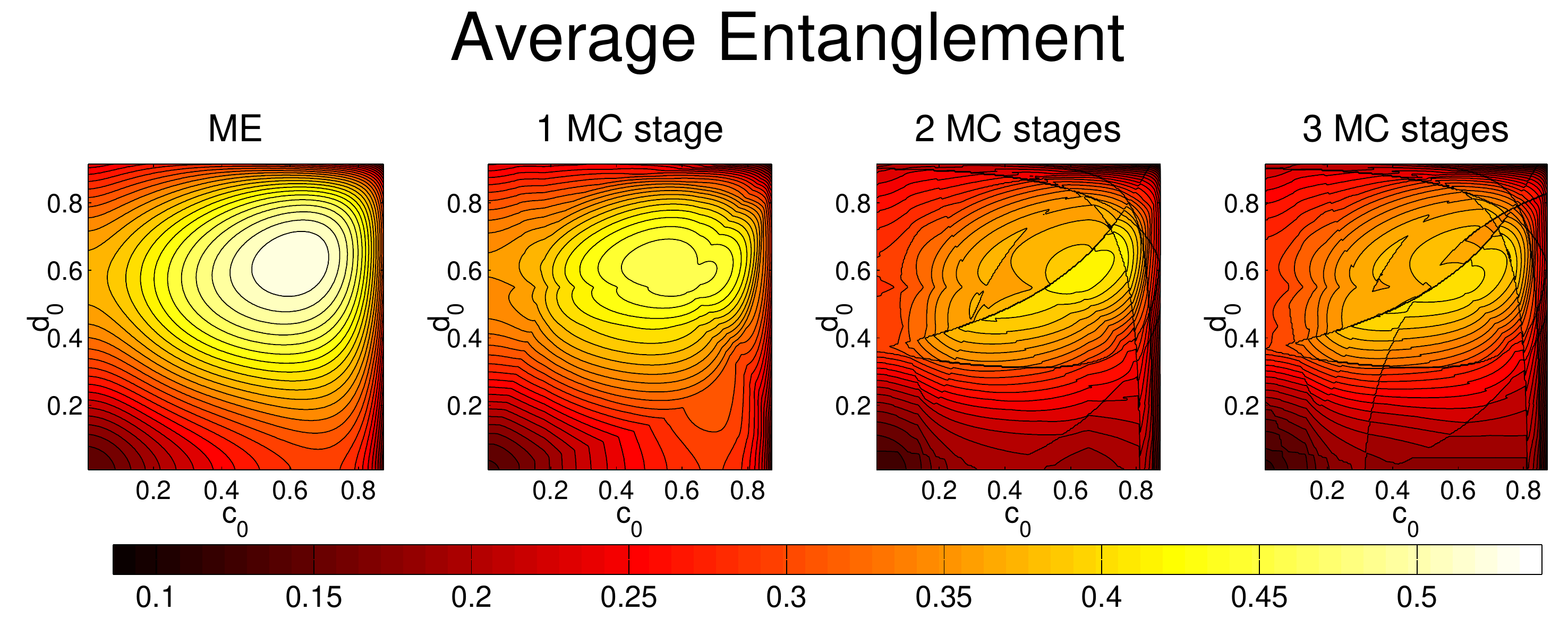}\vspace{0.5cm}
	\includegraphics[width=0.49\textwidth]{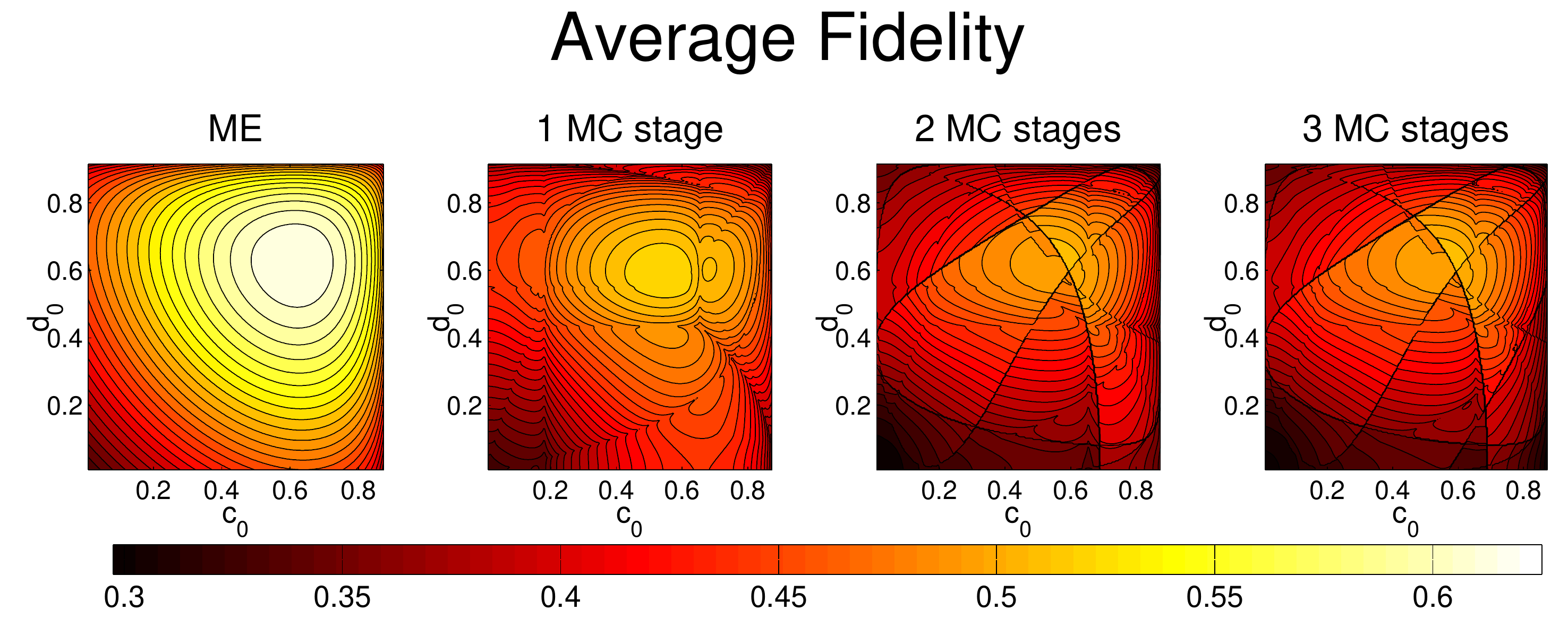}
	\caption{(Color online) A simulation of the overall average entanglement (upper panels) and fidelity (lower panels), considering every possible outcomes combination, when the maximum number of MC stages allowed at most is increased. Unlike the results shown in Fig.~\ref{fig:Contour_average}, the number of discrimination stages is chosen for every value of $s$ subject to $\beta_{s}^{\rm M}\leqslant 1, 2$, and $3$, respectively, in order to optimize the averages. The Schmidt coefficients are the same used for Figs.~\ref{fig:entandfid}--\ref{fig:Contour_average}. 
     \label{fig:Contour_adapt}}
\end{figure}

\begin{figure}[t!]
	\includegraphics[width=0.49\textwidth]{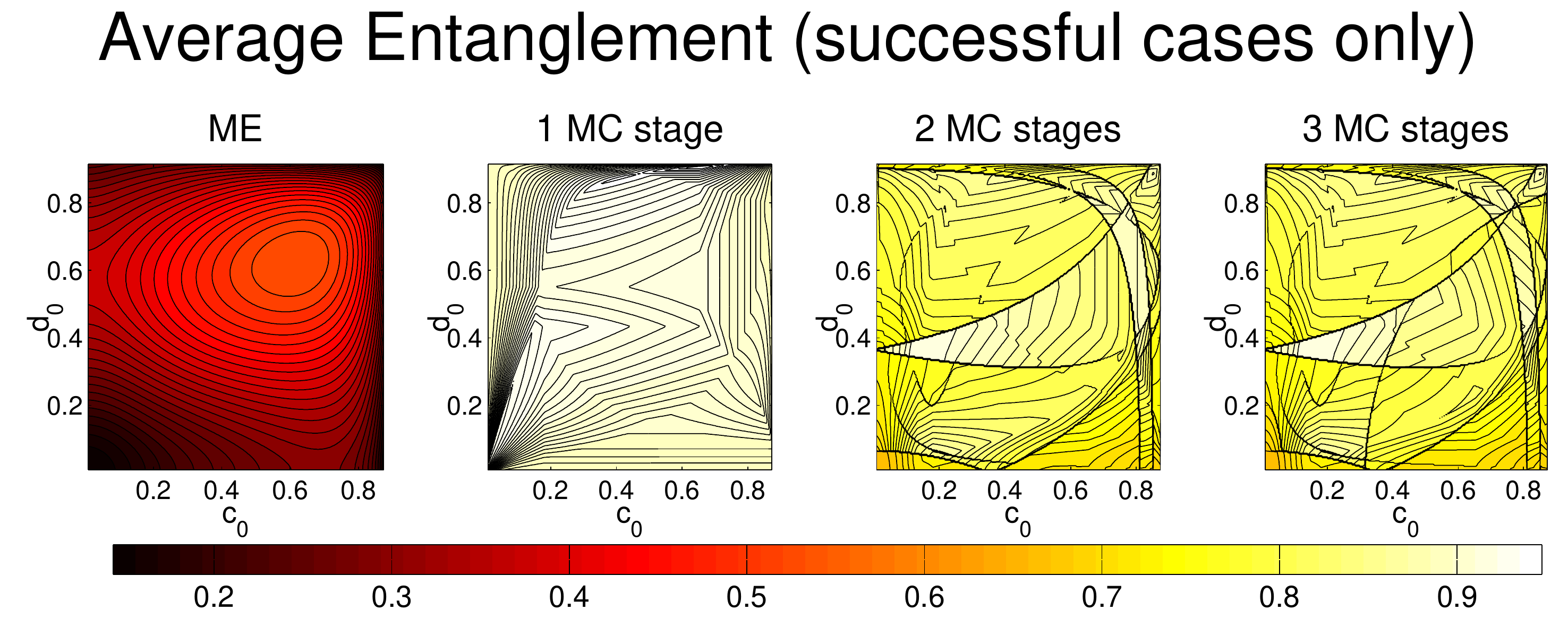}\vspace{0.5cm}
	\includegraphics[width=0.49\textwidth]{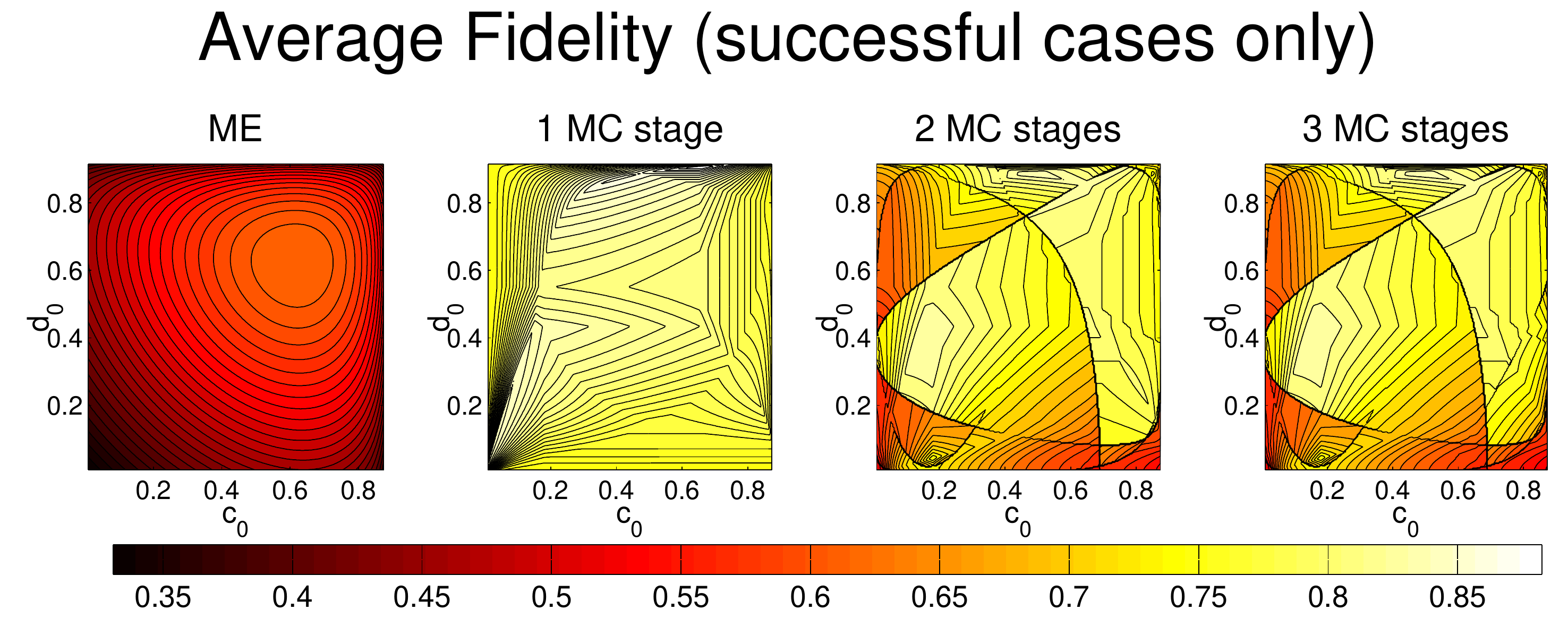}
	\caption{(Color online) A simulation of the overall average, concerning only the successful cases, of entanglement (upper panels) and fidelity (lower panels), considering every possible outcomes combination, when the maximum number of MC stages allowed at most is increased. As for the results shown in Fig.~\ref{fig:Contour_adapt}, the number of discrimination stages is set for every value of $s$ subject to $\beta_{s}^{\rm M}\leqslant 1, 2$, and $3$, respectively, optimizing the averaged figures of merit. The Schmidt coefficients are the same as those used for Figs.~\ref{fig:entandfid}--\ref{fig:Contour_adapt}. 
     \label{fig:Contour_succ_only}}
\end{figure}

\par In order to test this probabilistic protocol, numerical simulations have been carried out. The results are shown in Fig.~\ref{fig:entandfid}. We have considered two parties (${D_A=4}$ and ${D_B=5}$) whose Schmidt coefficients have been fixed as ${c_1 = 0.2811}$, ${c_2 = 0.3790}$, ${d_1 = 0.3220}$, ${d_2 = 0.2064}$ and ${d_3 = 0}$. The coefficients $c_0$ and $d_0$ are free and the normalization of the entangled states constrain the possible values of $c_3$ and $d_4$. It can be seen that, depending on the found value of $s$ and the features of the initial quantum channels, even a 3-stage MC implementation could lead to figures of merit surpassing the ones attainable by the standard protocol, as shown for ${s=1}$. For the other values of $s$, the third stage will lead to factorized states and no entanglement is transferred at all. The reason for this behavior is that the states $|\nu_{ls}\rangle$ for these cases are already linearly dependent. Unlike ${s=1}$ in which $\beta_{1}^{\rm M}=D_A-1$ stages can be set up for some Schmidt coefficients, for the other values of $s$ there exist fewer possible MC stages that worth implementing to obtain information about the states $|\nu_{ls}\rangle$. In these cases, however, a two-stage MC implementation can still lead to shared entangled states between Alice and Bob and, depending on the initial Schmidt coefficients, to probabilistically outperform the standard protocol. Note that the number of MC stages that surpass the performance of the standard procedure can vary, for the same Schmidt coefficients, depending on the value of $s$. Then, Charlie can use this class of information to determine the number $\beta_s^{\rm M}$ of necessary stages for each case. 

\par Besides having additional chances, the probability shown in Eq.~(\ref{prob_succbeta}) of swapping with improved figures of merit than the standard protocol can also be increased when we allow more MC stages to be implemented, as Fig.~\ref{fig:Contour_probability} shows. Note that, for this example, the probability of having the entanglement or the fidelity enhanced is either unchanged or increased according the number of stages increases too. 

\par However, it is worth mentioning that the improvements achievable via MC measurements are probabilistic. Indeed, the average entanglement and fidelity of Eqs.~(\ref{eq:MEentav}) and~(\ref{eq:MEfidav}), which correspond to ME discrimination, are higher than the averages of Eqs.~(\ref{eq:MCentav}) and~(\ref{eq:MCfidav}) obtained by SMC measurements when a fixed number $\beta_s^{\rm M}$ of stages are employed, as Fig.~\ref{fig:Contour_average} indicates considering $\beta_s^{\rm M}= 1,2,$ and $3$ as upper bounds. Additionally, if the number of stages $\beta_s^{\rm M}$ that worth implementing is adequately chosen for each possible value of $s$, the average values of entanglement and fidelity are larger than the ones of the case of fixed stages, as Fig.~\ref{fig:Contour_adapt} shows, but are still smaller than the average obtainable by means of the standard procedure. These results were expected because the largest average fidelity of transformation of a pure bipartite state into another using local operations and classical communication is achieved deterministically rather than probabilistically~\cite{Vidal00}. Such a behavior was also numerically shown in Ref.~\cite{Neves12} for quantum teleportation. Since the processes involved in ES are similar to the ones used in quantum teleportation, a similar nature about the average of both the entanglement and the fidelity was expected.

When we restrict ourselves to the successful cases only and when the number of stages $\beta_s^{\rm M}$ is adequately chosen for each attainable $s$, the average of the figures of merit increases due to the postselection that discards the unsuccessful cases. As Fig.~\ref{fig:Contour_succ_only} indicates, the averages are now larger than the ones of the standard procedure. However, according the number of stages increases, the averages decrease. This behavior shows a good agreement with the one exposed in the previous figures.

\section{Summary \label{sec:Summary}}
\par We have studied the entanglement swapping protocol for entangled states whose Schmidt number is not maximal. Depending on the experimental resources, a reduction to a smaller effective system in which the Schmidt number is maximal can be carried out or not. When it is not possible to do so, the protocol can contain the problem of discriminating among sets of linearly dependent states, for which two strategies were studied: ME and MC discrimination. When the aforementioned reduction is achievable, the states to be discriminated are linearly independent, MC reduces to UD, and both ME and UD are suitable. The latter strategy can be considered as a particular case of MC discrimination when it is applied on linearly independent states. An important difference between these strategies lies in the fact that ME is a deterministic operation and reproduces the operations performed in the standard ES protocol, whereas that MC and UD are probabilistic processes. Regardless which is the linear dependence of the states being handled, the use of the probabilistic strategy allows us to enhance some figures of merit such as entanglement and fidelity when a first stage of MC measurements is set up. Furthermore, when the probabilistic approach fails, additional attempts can be done to extract information. This is the core of sequential maximum confidence measurements. Depending of the Schmidt coefficients of the initial pairs of particles, it is possible to perform several stages of sequential maximum confidence measurements that outperform the standard protocol. Moreover, the existence of additional stages can increase the probability of enhancement of the protocol. However, these improvements are probabilistic: both entanglement and fidelity, when averaged over every possible measurement result, fall behind the averages obtained by ME measurements.

\begin{acknowledgments}
	This work was supported by CONICYT PFB08-24, Milenio ICM P10-030-F, and FONDECYT Grant No. 11121318. M.A.S.P. acknowledges the financial support from CONICYT.
\end{acknowledgments}


\begin{thebibliography}{99}

	\bibitem{Ladd10} T. D. Ladd, F. Jelezko, R. Laflamme, Y. Nakamura, C. Monroe, and J. L. O'Brien, \href{http://dx.doi.org/10.1038/nature08812}{Nature {\bf 464}, 45 (2010).}
	\bibitem{Kok07} P. Kok, W. J. Munro, K. Nemoto, T. C. Ralph, J. P. Dowling, and G. J. Milburn, \href{http://dx.doi.org/10.1103/RevModPhys.79.135}{\rmp {\bf 79}, 135 (2007).}
	\bibitem{Gisin02} N. Gisin, G. Ribordy, W. Tittel, and H. Zbinden, \href{http://dx.doi.org/10.1103/RevModPhys.74.145}{\rmp {\bf 74}, 145 (2002).}
	\bibitem{Jennewein00} T. Jennewein, C. Simon, G. Weihs, H. Weinfurter, and A. Zeilinger, \href{http://dx.doi.org/10.1103/PhysRevLett.84.4729}{\prl {\bf 84}, 4729 (2000).}
	\bibitem{Naik00} D. S. Naik, C. G. Peterson, A. G. White, A. J. Berglund, and P. G. Kwiat, \href{http://dx.doi.org/10.1103/PhysRevLett.84.4733}{\prl {\bf 84}, 4733 (2000).}
	\bibitem{Tittel00} W. Tittel, J. Brendel, H. Zbinden, and N. Gisin, \href{http://dx.doi.org/10.1103/PhysRevLett.84.4737}{\prl {\bf 84}, 4737 (2000).}
	\bibitem{Cleve97} R. Cleve and H. Buhrman, \href{http://dx.doi.org/10.1103/PhysRevA.56.1201}{\pra {\bf 56}, 1201 (1997).}

	\bibitem{Bennett93} C. H. Bennett, G. Brassard, C. Cr\'epeau, R. Jozsa, A. Peres, and W. K. Wootters, \href{http://dx.doi.org/10.1103/PhysRevLett.70.1895}{\prl {\bf 70}, 1895 (1993).}
	\bibitem{Zukowski93} M. \.Zukowski, A. Zeilinger, M. A. Horne, and A. K. Ekert, \href{http://dx.doi.org/10.1103/PhysRevLett.71.4287}{\prl {\bf 71}, 4287 (1993).}
	

	\bibitem{Sangouard11} N. Sangouard, C. Simon, H. de Riedmatten, and N. Gisin, \href{http://dx.doi.org/10.1103/RevModPhys.83.33}{\rmp {\bf 83}, 33 (2011).}
	
	\bibitem{Briegel98} H.-J. Briegel, W. D\"ur, J. I. Cirac, and P. Zoller, \href{http://dx.doi.org/10.1103/PhysRevLett.81.5932}{\prl {\bf 81}, 5932 (1998).}
	\bibitem {Dur99} W. D\"ur, H.-J. Briegel, J. I. Cirac, and P. Zoller, \href{http://dx.doi.org/10.1103/PhysRevA.59.169}{\pra {\bf 59}, 169 (1999).}
	\bibitem{Waks02} E. Waks, A. Zeevi, and Y. Yamamoto, \href{http://dx.doi.org/10.1103/PhysRevA.65.052310}{\pra {\bf 65}, 052310 (2002).}

	\bibitem{Shi00} B.-S. Shi, Y.-K. Jiang, and G.-C. Guo, \href{http://dx.doi.org/10.1103/PhysRevA.62.054301}{\pra {\bf 62}, 054301 (2000). }
	\bibitem{Hsu02} L.-Y. Hsu, \href{http://dx.doi.org/10.1016/S0375-9601(02)00297-9}{Phys. Lett. A {\bf 297}, 126 (2002).}
	\bibitem{Modlawska08} J. Mod{\l}awska and A. Grudka, \href{http://dx.doi.org/10.1103/PhysRevA.78.032321}{\pra {\bf 78}, 032321 (2008).}
	\bibitem{Yang09} M. Yang, A. Delgado, L. Roa, and C. Saavedra, \href{http://dx.doi.org/10.1016/j.optcom.2008.12.042}{Opt. Commun. {\bf 282}, 1482 (2009).}

	\bibitem{Bose98} S. Bose, V. Vedral, and P. L. Knight, \href{http://dx.doi.org/10.1103/PhysRevA.57.822}{\pra {\bf 57}, 822 (1998).}
	\bibitem{Hardy00} L. Hardy and D. D. Song, \href{http://dx.doi.org/10.1103/PhysRevA.62.052315}{\pra {\bf 62}, 052315 (2000).	}
	
	\bibitem{Zhou09} P. Zhou, F.-G. Deng, and H.-Y. Zhou, \href{http://dx.doi.org/10.1088/0031-8949/79/03/035005}{Phys. Scr. {\bf 79}, 035005 (2009).}
	\bibitem{Xia07} Y. Xia, J. Song, and H.-S. Song, \href{http://dx.doi.org/10.1088/0031-8949/76/4/014}{Phys. Scr. {\bf 76}, 363 (2007).}
	\bibitem{Zhan09} Y.-B. Zhan, L.-L. Zhang, and Q.-Y. Zhang, \href{http://dx.doi.org/10.1016/j.optcom.2009.08.024}{Opt. Commun. {\bf 282}, 4633 (2009).}
	\bibitem{Qin10}	S.-J. Qin, F. Gao, Q.-Y. Wen, and F.-C. Zhu, \href{http://dx.doi.org/10.1016/j.optcom.2009.11.087}{Opt. Commun. {\bf 283}, 1566 (2010).}
	\bibitem{Scherer11} A. Scherer, B. C. Sanders, and W. Tittel, \href{http://dx.doi.org/10.1364/OE.19.003004}{Opt. Exp. {\bf 19}, 3004 (2011).}
	\bibitem{Zhou11} N. Zhou, L. Wang, L. Gong, X. Zuo, and Y. Liu, \href{http://dx.doi.org/10.1016/j.optcom.2011.05.002}{Opt. Commun. {\bf 284}, 4836 (2011).}
	
	
	\bibitem{Boulant03} N. Boulant, K. Edmonds, J. Yang, M. A. Pravia, and D. G. Cory, \href{http://dx.doi.org/10.1103/PhysRevA.68.032305}{\pra {\bf 68}, 032305 (2003).}

	\bibitem{Riebe08} M. Riebe, T. Monz, K. Kim, A. S. Villar, P. Schindler, M. Chwalla, M. Heinrich, and R. Blatt, \href{http://dx.doi.org/10.1038/nphys1107}{Nat. Phys. {\bf 4}, 839 (2008).}

	\bibitem{Pan98} J.-W. Pan, D. Bouwmeester, H. Weinfurter, and A. Zeilinger, \href{http://dx.doi.org/10.1103/PhysRevLett.80.3891}{\prl {\bf 80}, 3891 (1998).}
	\bibitem{Jennewein01} T. Jennewein, G. Weihs, J.-W. Pan, and A. Zeilinger, \href{http://dx.doi.org/10.1103/PhysRevLett.88.017903}{\prl {\bf 88}, 017903 (2001).}
	\bibitem{Sciarrino02} F. Sciarrino, E. Lombardi, G. Milani, and F. DeMartini, \href{http://dx.doi.org/10.1103/PhysRevA.66.024309}{\pra {\bf 66}, 024309 (2002).}
	\bibitem{Kaltenbaek09} R. Kaltenbaek, R. Prevedel, M. Aspelmeyer, and A. Zeilinger, \href{http://dx.doi.org/10.1103/PhysRevA.79.040302}{\pra {\bf 79}, 040302 (2009).}
	\bibitem{Riedmatten05} H. de Riedmatten, I. Marcikic, J. A. W. van Houwelingen, W. Tittel, H. Zbinden, and N. Gisin, \href{http://dx.doi.org/10.1103/PhysRevA.71.050302}{\pra {\bf 71}, 050302 (2005).}
	\bibitem{Halder07} M. Halder, A. Beveratos, N. Gisin, V. Scarani, C. Simon, and H. Zbinden, \href{http://dx.doi.org/10.1038/nphys700}{Nature Phys. {\bf 3}, 692 (2007).}
	\bibitem{Takesue09} H. Takesue and B. Miquel, \href{http://dx.doi.org/10.1364/OE.17.010748}{Opt. Express {\bf 17}, 10748 (2009).	}
	\bibitem{Sangouard11exp} N. Sangouard, B. Sanguinetti, N. Curtz, N. Gisin, R. Thew, and H. Zbinden, \href{http://dx.doi.org/10.1103/PhysRevLett.106.120403}{\prl {\bf 106}, 120403 (2011).}
	\bibitem{Xue12} Y. Xue, A. Yoshizawa, and H. Tsuchida, \href{http://dx.doi.org/10.1103/PhysRevA.85.032337}{\pra {\bf 85}, 032337 (2012).}
	\bibitem{Peres00} A. Peres, \href{http://dx.doi.org/10.1080/09500340008244032}{J. Mod. Opt. {\bf 47}, 139 (2000).}
	\bibitem{Ma12} X.-S. Ma, S. Zotter, J. Kofler, R. Ursin, T. Jennewein, \v{C}. Brukner and A. Zeilinger, \href{http://dx.doi.org/10.1038/nphys2294}{Nat. Phys. {\bf 8} 479 (2012).}
	
	\bibitem{Goebel08} A. M. Goebel, C. Wagenknecht, Q. Zhang, Y.-A. Chen, K. Chen, J. Schmiedmayer, and J.-W. Pan, \href{http://dx.doi.org/10.1103/PhysRevLett.101.080403}{\prl {\bf 101}, 080403 (2008).}
	
	\bibitem{Lu09}	C.-Y. Lu, T. Yang, and J.-W. Pan, \href{http://dx.doi.org/10.1103/PhysRevLett.103.020501}{\prl {\bf 103}, 020501 (2009).}
	

	\bibitem{Bose99} S. Bose, V. Vedral, and P. L. Knight, \href{http://dx.doi.org/10.1103/PhysRevA.60.194}{\pra {\bf 60}, 194 (1999).}

	\bibitem{Delgado05} A. Delgado, L. Roa, J. C. Retamal, and C. Saavedra, \href{http://dx.doi.org/10.1103/PhysRevA.71.012303}{\pra {\bf 71}, 012303 (2005).	}
	\bibitem{Alber01} G. Alber, A. Delgado, N. Gisin and I. Jex, \href{http://dx.doi.org/10.1088/0305-4470/34/42/307}{ J. Phys. A {\bf 34}, 8821 (2001). }
	\bibitem{Alber00} G. Alber, A. Delgado, N. Gisin and I. Jex, \href{http://arXiv.org/abs/quant-ph/0008022}{arXiv:quant-ph/0008022v1 (2000).}
	
	\bibitem{Ivanovic87} I. D. Ivanovic, \href{http://dx.doi.org/10.1016/0375-9601(87)90222-2}{Phys. Lett. A \textbf{123}, 257 (1987).}
	\bibitem{Dieks88} D. Dieks, \href{http://dx.doi.org/10.1016/0375-9601(88)90840-7}{Phys. Lett. A \textbf{126}, 303 (1988). } 
	\bibitem{Peres88} A. Peres, \href{http://dx.doi.org/10.1016/0375-9601(88)91034-1}{Phys. Lett. A \textbf{128}, 19 (1988).}
	\bibitem{Jaeger95} G. Jaeger and A. Shimony, \href{http://dx.doi.org/10.1016/0375-9601(94)00919-G}{Phys. Lett. A \textbf{197}, 83 (1995).}
	\bibitem{Peres98} A. Peres and D. R. Terno, \href{http://dx.doi.org/10.1088/0305-4470/31/34/013}{J. Phys. A \textbf{31}, 7105 (1998).}
	\bibitem{Chefles98} A. Chefles, \href{http://dx.doi.org/10.1016/S0375-9601(98)00064-4}{Phys. Lett. A {\bf 239}, 339 (1998).}

	\bibitem{Croke06} S. Croke, E. Andersson, S. M. Barnett, C. R. Gilson and J. Jeffers, \href{http://dx.doi.org/10.1103/PhysRevLett.96.070401}{\prl \textbf{96}, 070401 (2006).}
	\bibitem{Jimenez11} O. Jim\'enez, M. A. Sol\'is-Prosser, A. Delgado, and L. Neves, \href{http://dx.doi.org/10.1103/PhysRevA.84.062315}{\pra {\bf 84}, 062315 (2011).}
	\bibitem{Herzog12} U. Herzog. \href{http://dx.doi.org/10.1103/PhysRevA.85.032312}{\pra {\bf 85}, 032312 (2012).}
	
	
	\bibitem{Brassard98} G. Brassard, S. L. Braunstein, and R. Cleve, \href{http://dx.doi.org/10.1016/S0167-2789(98)00043-8}{Physica D {\bf 120}, 43 (1998).}
	\bibitem{NielsenBook} M. A. Nielsen and I. L. Chuang, \textit{Quantum Computation and Quantum Information} (Cambridge University Press, Cambridge, 2000).
	
	\bibitem{Daboul03} J. Daboul, X. Wang and B. C. Sanders, \href{http://dx.doi.org/10.1088/0305-4470/36/10/312}{J. Phys. A: Math. Gen. {\bf 36}, 2525 (2003).}

	\bibitem{BergouReview} J. Bergou, \href{http://dx.doi.org/10.1080/09500340903477756}{J. Mod. Opt. {\bf 57}, 160 (2010).}
	\bibitem{BergouBook}   J. A. Bergou, U. Herzog, and M. Hillery, {\it Lecture Notes in Physics} (Springer, Berlin, 2004), Vol. 649, pp. 417--465.


	\bibitem{Holevo73} A. S. Holevo, \href{http://dx.doi.org/10.1016/0047-259X(73)90028-6}{J. Multivariate Anal. \textbf{3}, 337 (1973).}
	\bibitem{HelstromBook} C. W. Helstrom, \textit{Quantum Detection and Estimation Theory} (Academic Press, New York, 1976).
	\bibitem{Ban97} M. Ban, K. Kurokawa, R. Momose, and O. Hirota, \href{http://dx.doi.org/10.1007/BF02435921}{Int. J. Theor. Phys. {\bf 36}, 1269 (1997).}
	\bibitem{Barnett01} S. M. Barnett, \href{http://dx.doi.org/10.1103/PhysRevA.64.030303}{\pra {\bf 64}, 030303 (2001).}


	\bibitem{Vedral97} V. Vedral, M. B. Plenio, M. A. Rippin, and P. L. Knight, \href{http://dx.doi.org/10.1103/PhysRevLett.78.2275}{\prl {\bf 78}, 2275 (1997).}
	\bibitem{Rungta01} P. Rungta, V. Bu\v{z}ek, C. M. Caves, M. Hillery, and G. J. Milburn, \href{http://dx.doi.org/10.1103/PhysRevA.64.042315}{\pra {\bf 64}, 042315 (2001).}


	\bibitem{Neves12} L. Neves, M. A. Sol\'is-Prosser, A. Delgado and O. Jim\'enez, \href{http://dx.doi.org/10.1103/PhysRevA.85.062322}{\pra {\bf 85}, 062322 (2012).}
	\bibitem{SolisProsser13} M. A. Sol\'is-Prosser, O. Jim\'enez, L. Neves, and A. Delgado, \href{http://dx.doi.org/10.1088/0031-8949/2013/T153/014058}{Phys. Scr. {\bf 2013}, 014058 (2013).}
	\bibitem{Vidal00} G. Vidal, D. Jonathan, and M. A. Nielsen, \href{http://dx.doi.org/10.1103/PhysRevA.62.012304}{\pra {\bf 62}, 012304 (2000).}

	
	
\end{thebibliography}
\end{document}